\pdfoutput=1
\documentclass[submission,Phys]{SciPost}

\usepackage{amsmath,amssymb}
\usepackage{booktabs,multirow,graphicx,tabularx,mathtools,slashed}
\usepackage[maxfloats=100]{morefloats}
\usepackage{color,xcolor,braket,comment}
\usepackage[normalem]{ulem}
\usepackage{enumitem}
\usepackage{feynmp}

\newcommand{\schi}{s_{Z'}}
\newcommand{\cchi}{c_{Z'}}
\newcommand{\tchi}{t_{Z'}}
\newcommand{\sschi}{2 s_{Z'} c_{Z'}}

\newcommand{\qqquad}{\qquad \qquad}
\newcommand{\qqqquad}{\qquad \qquad \qquad}
\setlist[itemize]{itemsep=1pt,parsep=1pt, topsep=3pt}
\setlist[enumerate]{itemsep=1pt,parsep=1pt, topsep=3pt}

\newcommand{\eg}{e.\,g.\ }
\newcommand{\met}{\slashchar{E}_T}

\newcommand{\ifb}{\ensuremath \mathrm{fb}^{-1}}
\newcommand{\iab}{\ensuremath \mathrm{ab}^{-1}}
\newcommand{\gev}{{\ensuremath \mathrm{GeV}}}
\newcommand{\tev}{{\ensuremath \mathrm{TeV}}}

\DeclareMathOperator{\br}{BR}

\newcommand{\arxiv}[1]{\href{http://arxiv.org/abs/#1}{arXiv:#1}}

\AtBeginDocument{
  \heavyrulewidth=.08em
  \lightrulewidth=.05em
  \cmidrulewidth=.03em
  \belowrulesep=.65ex
  \belowbottomsep=0pt
  \aboverulesep=.4ex
  \abovetopsep=0pt
  \cmidrulesep=\doublerulesep
  \cmidrulekern=.5em
  \defaultaddspace= .5em
  \setlength{\tabcolsep}{0.5em}
}

\def\slashchar#1{\setbox0=\hbox{$#1$}           
   \dimen0=\wd0                                 
   \setbox1=\hbox{/} \dimen1=\wd1               
   \ifdim\dimen0>\dimen1                        
      \rlap{\hbox to \dimen0{\hfil/\hfil}}      
      #1                                        
   \else                                        
      \rlap{\hbox to \dimen1{\hfil$#1$\hfil}}   
      /                                         
   \fi}

\begin{document}
\begin{fmffile}{feynman}

\begin{center}{\Large \textbf{
Dark Matter in Anomaly-Free Gauge Extensions
}}\end{center}

\begin{center}
Martin Bauer\textsuperscript{1},
Sascha Diefenbacher\textsuperscript{1},
Tilman Plehn\textsuperscript{1},
Michael Russell\textsuperscript{1}, \\ and
Daniel A. Camargo\textsuperscript{2}
\end{center}

\begin{center}
{\bf 1} Institut f\"ur Theoretische Physik, Universit\"at Heidelberg, Germany\\
{\bf 2} International Institute of Physics, Universidade Federal do Rio Grande do Norte, Campus Universitario, Lagoa Nova, Natal, Brazil \\
plehn@uni-heidelberg.de
\end{center}

\vspace{-.6cm}
\section*{Abstract}
{\bf A consistent model for vector mediators to dark matter needs to
  be anomaly-free and include a scalar mode from mass generation. For
  the leading U(1) extensions we review the structure and constraints,
  including kinetic mixing at loop level. The thermal relic density
  suggests that the vector and scalar masses are similar. For the LHC
  we combine a $\boldsymbol{Z'}$ shape analysis with mono-jets. For the latter, we
  find that a shape analysis offers significant improvement over
  existing cut-and-count approaches. Direct
  detection limits strongly constrain the kinetic mixing angle and we propose a $\boldsymbol{\ell^+\ell^-\met}$ search strategy based on the scalar mediator.}

\tableofcontents

\clearpage
\section{Introduction}
\label{sec:intro}

The nature of dark matter is one of the great mysteries in particle
physics and cosmology. A comprehensive experimental program is on the
way to identify the dark matter agent and determine its properties. On
the theory side, many years of intense research have convinced us that
perturbative gauge theories are the appropriate framework to describe
physics above the QCD scale. The leading dark matter candidate around
the weak scale is thermal freeze-out dark
matter~\cite{freezeout,reviews,lecture}, naturally predicting the
observed relic density for weak-scale or TeV-scale masses and
electroweak-sized couplings. \bigskip

The Standard Model allows for three renormalizable couplings to a mediator to the dark sector: The Higgs portal, the neutrino portal and the vector portal. The vector portal predicts a new spin-1 $Z'$ boson that couples to SM matter through a kinetic mixing term and can be searched for at colliders~\cite{zprime,u1_mediator,other_zprime}. For order one gauge couplings,  
%
%
%
%
the approximate relation $\Omega_\chi h^2
\approx 5 \cdot 10^{-10} \, \gev^{-2}/ \langle
\sigma_{\chi \chi} v\rangle$ relates the observed relic density to a
(large) dark matter annihilation rate. For example in the case of
$m_{Z'} = m_\chi~...~2 m_\chi$ this turns into the condition 
\begin{align}
\langle \sigma_{\chi \chi} v\rangle \approx \frac{g^4 m_\chi^2}{16 \pi m_{Z'}^4}
\quad \Rightarrow \quad 
\frac{m_{Z'}}{g^2} < 1~\tev \; .
\end{align}
with a dark matter mass $m_\chi$, a mediator mass $m_{Z'}$, and a
perturbative coupling $g$.  Similarly, for even heavier mediators with
on-shell decays to the dark matter agent, $m_{Z'} > 2 m_\chi$, we find
\begin{align}
\langle \sigma_{\chi \chi} v\rangle \approx \frac{g^4}{16 \pi m_{Z'}^2}
\quad \Rightarrow \quad 
\frac{m_{Z'}}{g^2} < 2~\tev \; .
\end{align}
This mediator mass range implies that a global analysis of thermally
produced dark matter with the observed relic density is described by a
fully propagating mediator at the LHC~\cite{eft2,simplified}.\bigskip

Besides a mediator that only communicates with Standard Model matter through kinetic mixing, SM particles could be gauged under the new gauge group. In this case a consistent ultraviolet complete model requires possible gauge anomalies to cancel.
  ~\cite{anomalies,anomalies_review}, for the generators of the gauge
group and a sum or trace over the relevant left-handed fermions.  If the $Z'$ mass is generated by a Higg-mechanism, the corresponding scalar can play an important role in phenomenology. 
Such a scalar is usually omitted in simplified models, in spite of
the absence of any formally applicable decoupling
argument~\cite{appelquist}.  Following both these arguments, an appropriate
simplified model of a heavy spin-1 mediator includes
\begin{itemize}
\item a gauge boson and a scalar describing the massive
  mediator sector~\cite{scalar};
\item either new fermions or an anomaly-free gauge
  group~\cite{anomaly_orig,anomaly_free}.
\end{itemize}
Both of these aspects need to be considered when we construct
meaningful models for dark matter with vector mediators. 

\bigskip

 
Anomaly-free gauge groups therefore provide well motivated mediators to dark matter. More general models are possible, but predict
 %
a sizable number of new fermions to cancel the anomalies, which contribute to 
%
interactions between the dark sector and the Standard Model~\cite{anomalies}.
%
%
We focus on anomaly-free gauge mediators based on the three
possible setups:
\begin{enumerate}
\item We assign all SM fermions as singlets under the new group
  $U(1)_X$ and only charge the dark matter fermion, which in turn does
  not couple to the SM gauge bosons. This setup is trivially free
  of anomalies, and the $Z'$ couplings to Standard Model fermions
  arise through kinetic mixing~\cite{mixing_orig,u1x_zprime}.

\item We choose an anomaly-free gauge group based on lepton number and
  utilize more than one generation for the anomaly condition.  Viable
  examples are the charged lepton number differences
  $U(1)_{L_\tau-L_\mu}, U(1)_{L_\tau-L_e}$, or $
  U(1)_{L_\mu-L_e}$~\cite{mu_tau_zprime}. Such models can be motivated
  for example through neutrino masses~\cite{mu_tau_zprime_neut} or
  flavor anomalies~\cite{mu_tau_zprime_b}.  The corresponding
  baryon-number-based constructions are usually ruled out by the
  observed structure of the CKM matrix.

\item We gauge the difference between baryon and lepton number
  $U(1)_{B-L}$~\cite{bminusl_zprime}. It has the specific advantage of
  allowing for Majorana masses for right-handed neutrinos after
  symmetry breaking~\cite{neutrino_mass}. In that sense, an
  anomaly-free $U(1)_{B-L}$ gauge group is motivated by a structural
  deficit of the Standard Model, because it requires right-handed
  leptons at some scale.
\end{enumerate}
We argue that searches for missing energy signals at the LHC are particularly powerful for two of these models, namely the $U(1)_X$ and the $U(1)_{L_\mu-L_\tau}$ gauge groups. After deriving the properties of the mediators for the three classes of models defined above, we focus our analysis on these two models. Other known anomaly-free $U(1)$ extensions include c
$U(1)_{L_\mu+L_\tau-2L_e}$ or $U(1)_R$, where right-handed SM fields
carry charges proportional to $T^3$ of $SU(2)_R$. However, their
phenomenology is not expected to be fundamentally different from the
three above cases, and in some cases the structure is actually
equivalent~\cite{ennio}.\bigskip

In this paper we will first introduce a kinetically mixed gauge
extension and the Higgs-like scalar in Sec.~\ref{sec:general} and  discuss the three anomaly-free gauge extensions we focus on in the remainder of this paper. This includes not only the general structure
of the model, but also the decay modes of the heavy gauge bosons and
their Higgs-like scalars. In Sec.~\ref{sec:constraints} we will
collect all available constraints from low-energy and collider
data. The properties of the new particle as a dark matter mediator will
be the focus of Sec.~\ref{sec:dm}. Finally, we will compare different
LHC strategies for searching for the new heavy states in
Sec.~\ref{sec:lhc}.

\section{U(1)-gauge extensions}
\label{sec:general}

We consider consistent dark matter models with a spin-1
mediator $Z'$ and a dark matter fermion $\chi$, charged under the new
gauge group. The available options are purely singlet SM fermions,
gauged lepton number differences, or the well-known anomaly-free
difference between the lepton and baryon numbers~\cite{patrick},
\begin{align}
U(1)_X\,, \qqquad 
U(1)_{L_i-L_j} \,,\qqquad 
U(1)_{B-L} \; ,
\label{eq:gg}
\end{align}
with $i\neq j=1,2,3$. The $Z'$ couplings to currents of SM fermions
are given by
\begin{alignat}{9}
\mathcal{L}_\text{fermion} = -g_{Z'} j'_\mu & {Z'}^\mu \notag \\
j'_\mu&= 0 \qquad && U(1)_X \notag  \\
j'_\mu&= \bar L_i \gamma_\mu L_i 
          + \bar \ell_i\gamma_\mu \ell_i 
          - \bar L_j \gamma_\mu L_j -\bar\ell_j\gamma_\mu \ell_j
            \qquad && U(1)_{L_i-L_j} \notag \\ 
j'_\mu&=  \frac{1}{3}\bar Q \gamma_\mu Q 
          + \frac{1}{3}\bar u_R\gamma_\mu u_R 
          + \frac{1}{3}\bar d_R\gamma_\mu d_R
          - \bar L \gamma_\mu L 
          + \bar \ell\gamma_\mu \ell
            \qquad && U(1)_{B-L} \; ,
\label{eq:fermion_currents}
\end{alignat}
where $g_{Z'}$ denotes the dark gauge coupling.  The different
coupling structures shown above can be understood in terms of a flavor
structure of a dark gauge coupling matrix. 

The fermion current structure of Eq.\eqref{eq:fermion_currents} can be
generalized to include the dark matter current. To couple to the gauge
mediator the dark matter fermion has to be a Dirac fermion. To avoid
new anomalies, the dark matter candidate cannot be chiral and its
charges under the new gauge group are $q_{\chi_L}=q_{\chi_R}$. This
defines a dark fermion Lagrangian with a vector mass term
\begin{align}
\mathcal{L}_\text{DM}= i \bar \chi \slashed D \chi - m_\chi \bar \chi \chi \; ,
\end{align}
with the covariant derivative of the SM-singlet fermion $D_\mu
=\partial_\mu -ig_{Z'} q_\chi \hat Z'_\mu$.\bigskip

In all cases, the kinetic term for the $U(1)$ gauge bosons is not canonically normalized

\begin{align}
\mathcal{L}_\text{gauge} 
= -\frac{1}{4}
\begin{pmatrix} \hat{B}_{\mu \nu} & \hat{Z}'_{\mu \nu} \end{pmatrix}
\begin{pmatrix} 1 & \schi \\ \schi & 1 \end{pmatrix}
\begin{pmatrix} \hat{B}_{\mu \nu} \\ \hat{Z}'_{\mu \nu} \end{pmatrix} \; ,
\label{eq:kinmixlag}
\end{align}
and afternormalizing the kinetic terms and rotating to the mass eigenbasis, the masses of the vector bosons are given by
\begin{align}
m_\gamma &= 0 \notag \\
m_{Z}^2&=
 \dfrac{v^2}{4}(g^2+{g'}^2) \; \left(1-\dfrac{v^2}{v_S^2} \; \dfrac{\schi^2{g'}^2}{8g_{Z'}^2 q_S^2}\right)
 + \mathcal{O}\left( \dfrac{v^6}{v_{S}^4} \right) \\
 m_{Z'}^2&=
 \dfrac{g_{Z'}^2q_S^2v_S^2}{2\cchi^2} + \dfrac{v^2}{4}{g'}^2\tchi^2
 + \mathcal{O}\left( \dfrac{v^4}{v_S^2} \right)  \;.
\end{align}
For details of the calculation, we refer the reader to Appendix~\ref{app:details}. 

As a second structural ingredient we give mass to the new gauge boson
by introducing a complex scalar $S$ with the potential
\begin{align}\label{eq:scalar}
\mathcal{L}_\text{scalar}
= \frac{1}{2}\, ( D_\mu S) (D^\mu S)^\dagger
 + \mu_S^2 \, S^\dagger S 
 + \frac{\lambda_S}{2} (S^\dagger S)^2 
 + \lambda_{HS} \, H^\dagger H \, S^\dagger S\; .
\end{align}
In this case the covariant derivative introduces the charge $q_S$ of
the heavy scalar under the new gauge group.  \bigskip

Under the conservative assumption that
SM gauge couplings and the Higgs vacuum expectation value are fixed,
the error $m_Z=91.1876\pm 0.0021$ \cite{Patrignani:2016xqp} constrains the mixing to
\begin{align}
\frac{g_{Z'}q_S}{\schi} \; v_S \gtrsim 1.3 \,\tev \qquad \text{at}\quad 95\%\, \text{CL}\; .
\label{eq:zmassconstraint}
\end{align}
It is interesting to compare the mass parameters for the heavy new
scalar and the heavy new vector modes in the mass matrices of
Eq.\eqref{eq:matrix_scalars} and Eq.\eqref{eq:matrix_vectors}
\begin{align}
\frac{m_S}{m_{Z'}} 
\sim \frac{\sqrt{\lambda_S}}{g_{Z'} q_S/\cchi}\,,
\label{eq:mass_ratio}
\end{align}
where we identify the heavy entries in the mass matrices with the new
masses and ignore parameters which are expected to be of order
one. Separating these two mass scales is not impossible, but requires
a dedicated model building effort, which means that a generic analysis
of gauge extensions should include the scalar mode in the mediator
sector.\bigskip

The couplings of the mass eigenstates to fermions and scalars play an important role in the following analysis and we find
\begin{align}
\mathcal{L_\text{fermion}}&= ej_\text{em} A \notag\\
&\phantom{=}- c_w s_3 t_{Z'} ej_\text{em} Z  +(c_3+s_ws_3t_{Z'})\frac{e}{s_wc_w}j_Z Z  + \frac{s_3}{c_{Z'}}g_{Z'}j_{Z'} Z\notag\\
&\phantom{=}- c_w c_3 t_{Z'} ej_\text{em} Z' +(s_wc_3t_{Z'}-s_3)\frac{e}{s_wc_w}j_Z Z'
  + \frac{c_3}{c_{Z'}}g_{Z'}j_{Z'} Z'
\end{align}
and
\begin{align}
\mathcal{L_\text{scalar}}&\ni \frac{v}{8}(g^2+g^{\prime 2}) (c_\alpha H-s_\alpha S) Z_{\mu}Z^\mu \\
&\phantom{\ni}  +\frac{v}{4}s_wt_{Z'}(g^2+g^{\prime 2}) (c_\alpha H-s_\alpha S) Z_\mu Z^{\prime \mu}\notag\\
&\phantom{\ni}  +\frac{v}{8}  s_w^2 t_{Z'}^2\bigg[c_\alpha \bigg(g^2\!+\!g^{\prime 2}\!+\!\frac{4g_{Z'}^2 q_S^2 t_\alpha}{s_w^2s_{Z'}^2}\frac{v_S}{v} \bigg) H- s_\alpha  \bigg(g^2\!+\!g^{\prime 2}\!-\!\frac{4g_{Z'}^2 q_S^2 t_\alpha}{s_w^2 s_{Z'}^2}\frac{v_S}{v} \bigg)S\bigg] Z'_\mu Z^{\prime \mu}\notag\,.
\end{align}
%


The phenomenology of
anomaly-free $U(1)$-extensions can thus be described by a small number of
model parameters. The Lagrangian features the most relevant new
parameters
\begin{align}
\{ \; m_\chi, \, g_{Z'}, m_{Z'}, \schi, \, m_S, \lambda_{HS}\; \} \; .
\end{align}
The charges under the new $U(1)$-symmetry we assume to be of order
one. As long as we focus on a heavy dark matter mediator with
on-shell decays, $m_{Z'} > 2 m_\chi$, the dark matter mass mainly
enters the computation of the mediator widths $\Gamma_{S,Z'}$.\bigskip

The vector and scalar mediator masses are typically related, as shown in
Eq.\eqref{eq:mass_ratio}. A hierarchy with a comparably light scalar
$\lambda_S \ll g_{Z'}$ is possible, but not the focus of our
paper. Alternatively, the scalar can be heavier than the
vector, $g_{Z'}\ll \lambda_S< 4\pi$. In this case, the small gauge
coupling suppresses the interaction of the new gauge boson with the
Standard Model. This does not only affect the LHC production cross section, it
also reduces the annihilation cross section in the early universe to
the point where an efficient annihilation is only possible around the
pole condition $m_{Z'} = 2 m_\chi$.

The phenomenology of the vector mediator is determined by its
couplings to the Standard Model and by its mass $m_{Z'}$.  In
Eq.\eqref{eq:all_mixings} we see that couplings to SM fermions can
arise through kinetic mixing ($\tchi$), through mixing with the
$Z$-boson ($s_3$), or through the $U(1)$ charges of the fermions
($g_{Z'}$).

The properties of the new scalar $S$ are largely independent of the
dark matter properties. All couplings to a pair of SM particles
proceed through the Higgs portal ($s_\alpha$), with the possible
exception of a the coupling to right-handed neutrinos in the case of
$U(1)_{B-L}$. Interesting features only arise in couplings linking
both mediators, like the $Z'$-$S$-$Z$ coupling.

\subsection{$U(1)_X$}

\begin{figure}[b!]
\includegraphics[width=0.50\textwidth]{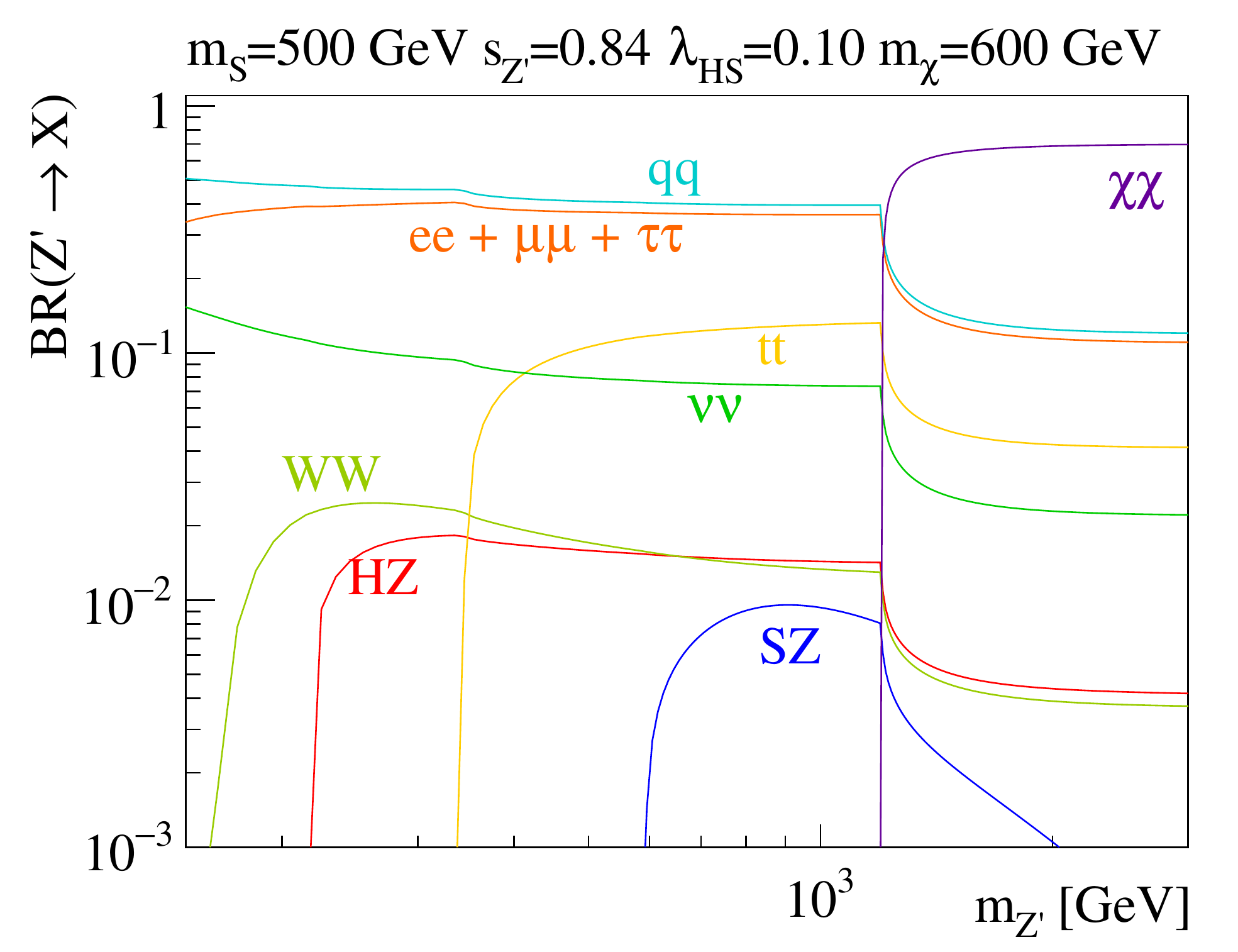}
\includegraphics[width=0.50\textwidth]{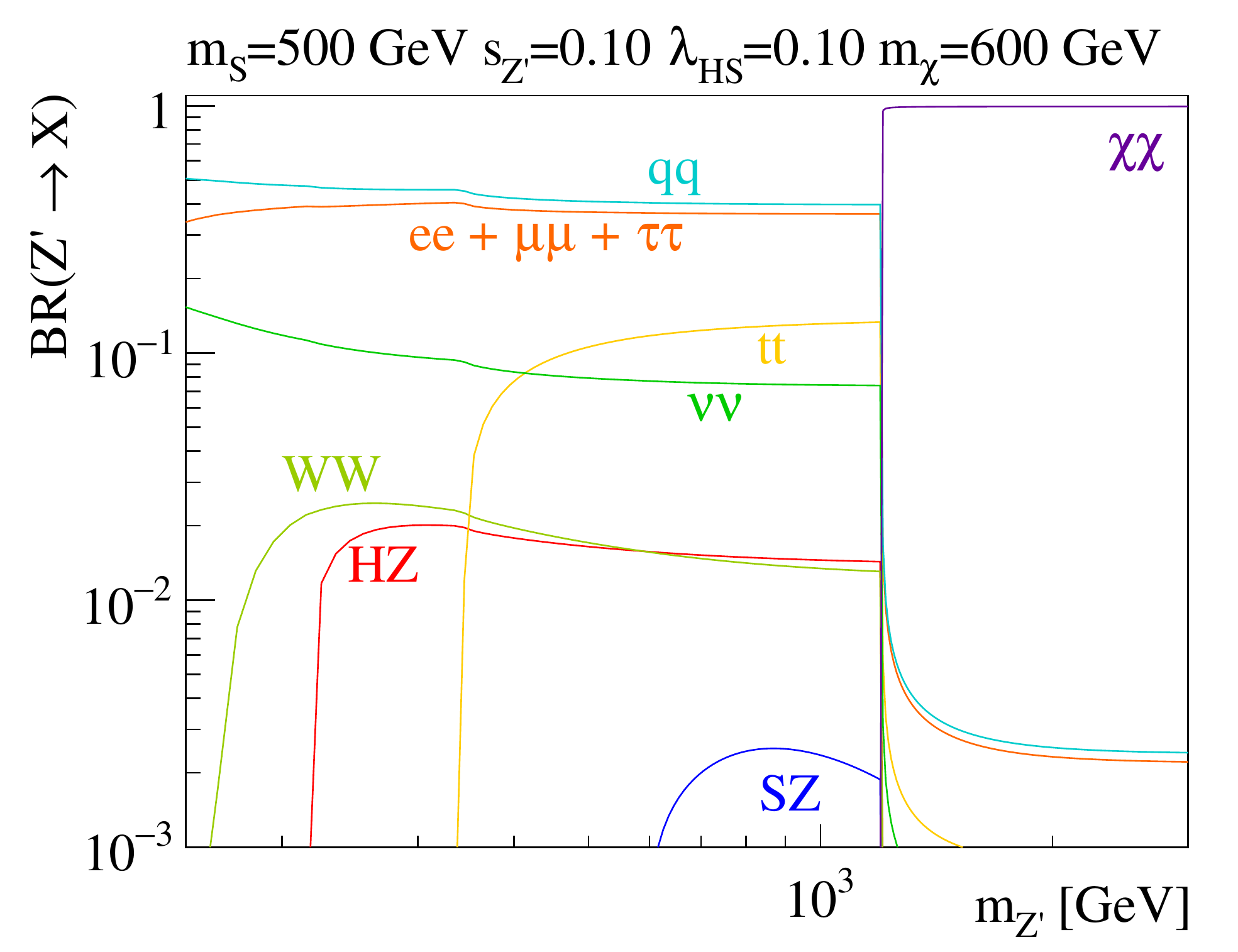}
\caption{Branching ratios for the $U(1)_X$ gauge boson with
  $m_S=500$~GeV, $\lambda_{HS}=0.1$, $g_{Z'}=1$, and $\schi = 0.84$
  (left) and $\schi = 0.1$ (right). The variable $m_{Z'}$ is varied
  through $v_S =50-1150$~GeV in the left panel and $v_S =
  100-2110$~GeV in the right panel. Correspondingly, the Higgs
  mixing angle varies between $s_\alpha=0.001-0.12$ (left) and $s_\alpha=0.008-0.23$.}
\label{fig:U1XBRs}
\end{figure}

\begin{figure}[t]
\includegraphics[width=0.50\textwidth]{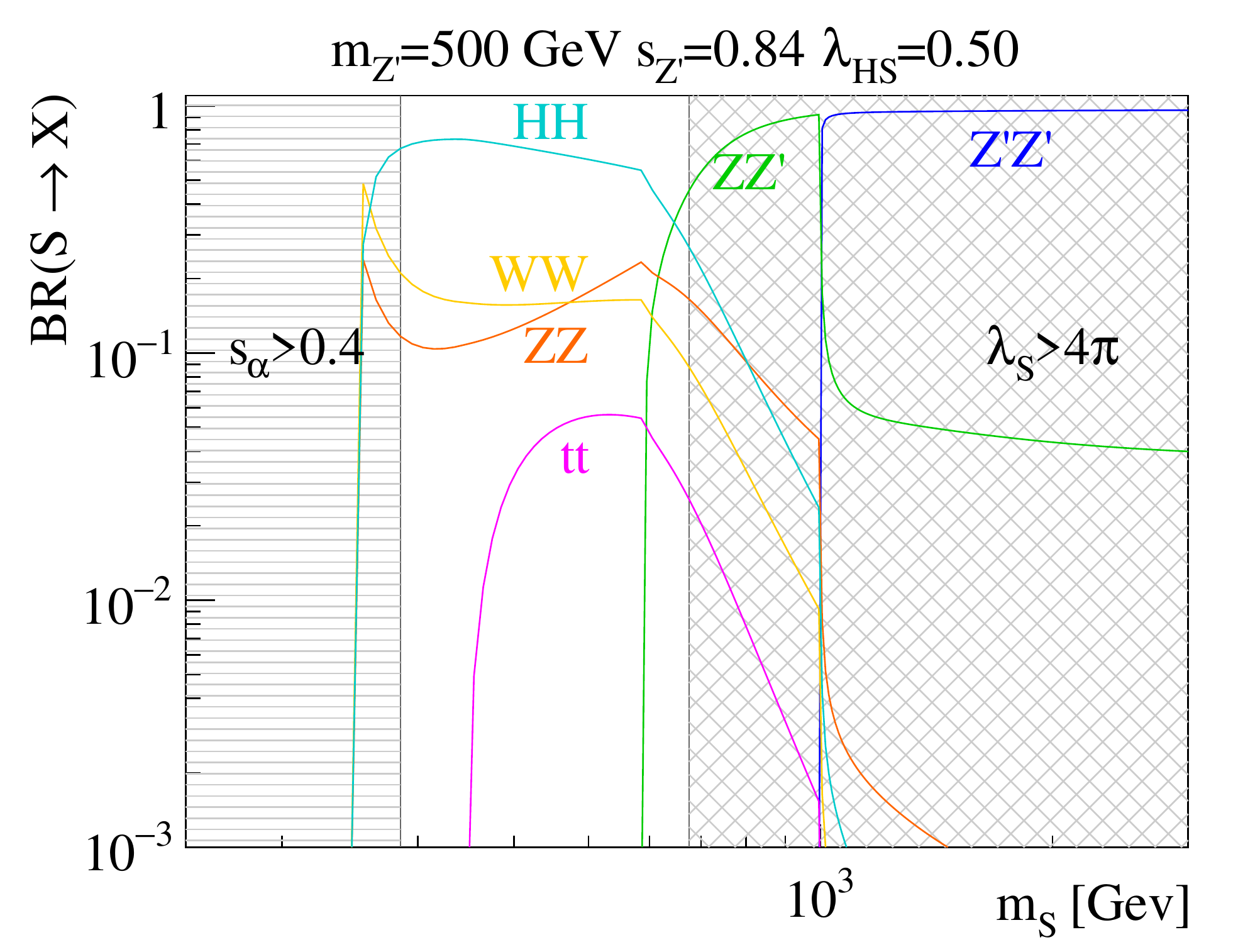}
\includegraphics[width=0.50\textwidth]{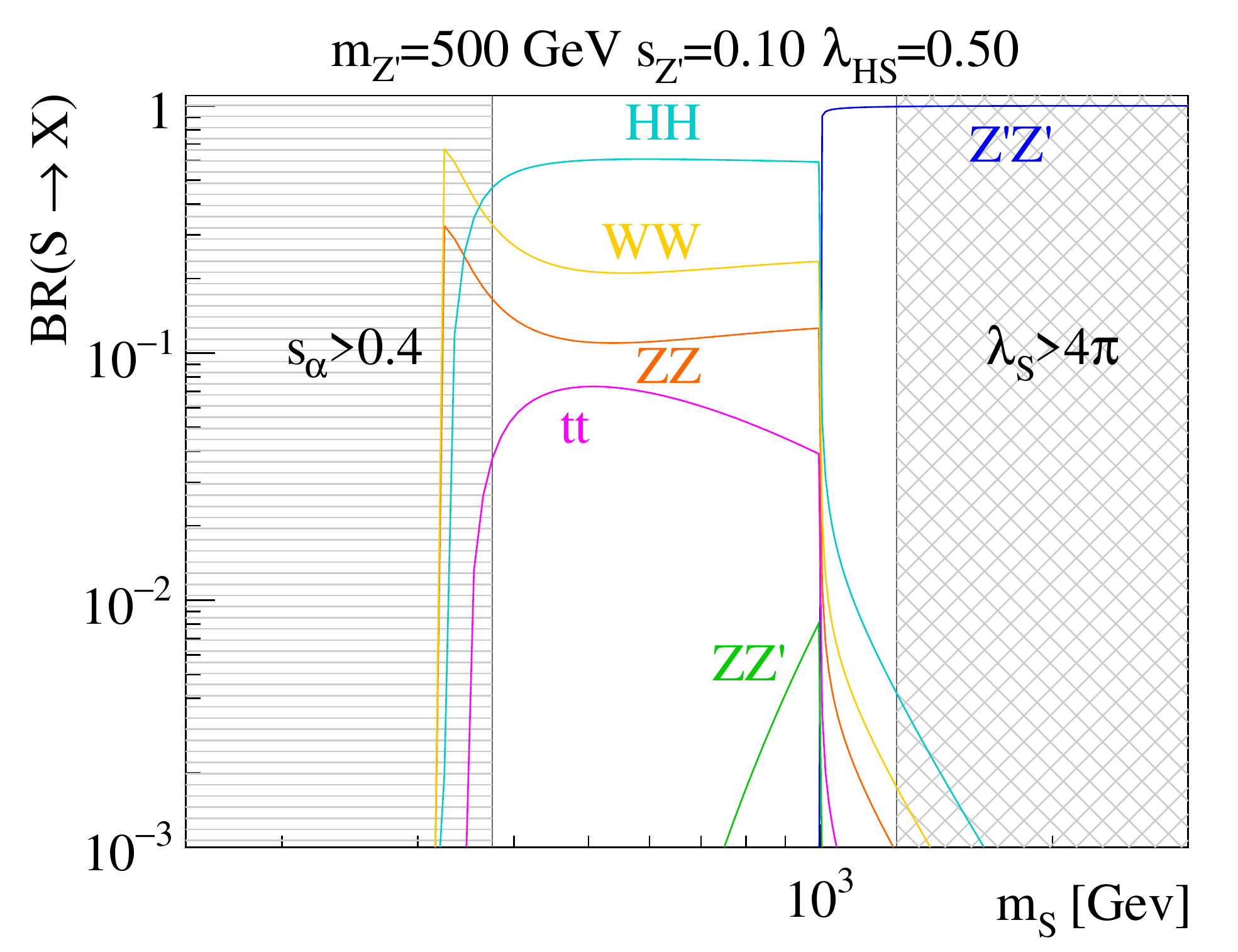}
\caption{Branching ratios for the heavy scalar $S$ with
  $m_{Z'}=500$~GeV, $\chi=1$ and $g_{Z'}=1$ and for different values
  of the quartic coupling $\lambda_{HS}$. The hashed regions are excluded by Higgs signal strengths measurements $(s_\alpha > 0.4)$ and perturbativity of the scalar potential $(\lambda_S > 4\pi)$.}
\label{fig:SBRs}
\end{figure}

In our first setup all SM particles are singlets under the new $U(1)$
gauge symmetry and all $Z'$ and $S$ couplings to the Standard Model
are induced by mixing. Both of these mixing effects lead to a
transition between the SM-sector and the dark matter sector of the
theory. The relation between the kinetic mixing in the gauge sector
and the scalar mixing in the Higgs sector reflects a symmetry
structure reminiscent of gaugino and scalar masses in broken
supersymmetry~\cite{susy_mass}. While the gauge-kinetic mixing is
protected by the additional gauge coupling and only multiplicatively
renormalized, the scalar mixing is just a property of the Higgs
potential. This is why quantum effects transform gauge-kinetic mixing
into a finite scalar mixing, but scalar mixing does not induce mixing
in the gauge sector.\bigskip

As a starting point, we show the $Z'$ branching ratios in
Fig.~\ref{fig:U1XBRs}, assuming two values of sizeable kinetic
mixing. Clearly, $Z'$ decays to two dark matter fermions through an
un-suppressed $U(1)_X$ charge dominates, provided the process $Z' \to
\chi \bar{\chi}$ is kinematically allowed. The partial widths to SM
fermions are universally proportional to $\schi$, including the $Z'
\to \nu \bar{\nu}$ background to the dark matter signal.  Due to the
non-orthogonal mixing in Eq.\eqref{eq:all_mixings} the electromagnetic
current contributes, so the structure of the $Z'$ branching ratios
does not correspond to $Z$-decay channels. Decays to light quark pairs
reach branching ratios of $30\%~...~40\%$, enhanced by color
factors. Decays to leptons amount to almost the same rate. The
$t\bar{t}$ decay channel exceeds 10\% slightly above its threshold.

The bosonic decays $Z'\to S Z, HZ$ can reach per-cent-level branching
ratios.  Other bosonic channels, like $Z' \to ZZ$ or $Z' \to HH$ are
not possible. For the two dominating bosonic channels we find that the
leading diagrams lead to a scaling
\begin{align}
\frac{\br (Z'\to SZ)}{\br(Z'\to HZ)} 
\propto  t_\alpha^2\; , 
\end{align}
with the larger $\br(Z'\to ZH)$ at the per-cent level. The
mixing angle $s_\alpha$ changes for the different values of $\schi$
used in Figs.~\ref{fig:U1XBRs} and ~\ref{fig:ULmutauXBRs}, because
$v_S$ depends on this choice when all other parameters remain the
same.  While one might expect effects from mass insertions $1/m_Z$ or
$1/m_{Z'}$ in this ratio, the corresponding diagrams for the decay $Z'
\to SZ$ are sub-leading for finite Higgs mixing. As we will see, for
sizable kinetic mixing the decay $Z'\to HZ$ combined with a large
invisible decay rate provides a tell-tale signal of this type of
models.\bigskip

For small kinetic mixing, the only way to produce the $Z'$ with a
sizeable rate is $S$-production with a decay $S \to Z'Z'$.  The
branching ratios of the heavy scalar $S$ for small and large gauge
mixing $\schi$ are shown in Fig.~\ref{fig:SBRs}. We indicate
constraints by perturbativity, $\lambda_S > 4\pi$, and global Higgs
analysis results, $s_\alpha< 0.4$ at 68\% C.L.~\cite{legacy}. Decays
to the SM Higgs boson or SM gauge bosons, mediated by the Higgs
portal, dominate over a wide range of parameters.  The mixed decay $S
\to Z Z'$ turns on for large kinetic mixing $\schi$, but for both
choices of $\schi$ the direct decay to $Z'Z'$ pairs completely
dominates once it is allowed. The only caveat is that in this regime
the self-coupling $\lambda_S$, responsible for the mass of the heavy
scalar, can become very large.

Finally, looking at the models there exists a fundamental difference
between the kinetic gauge mixing and the Higgs mixing.  If the
$U(1)_X$ group is embedded in a non-abelian gauge group $SU(N)_X$ at a
higher scale, kinetic mixing is never generated. On the other hand no
symmetry principle forbids a Higgs portal.  In this limit, our
$U(1)_X$ model corresponds to a Higgs-portal model with an dark sector
consisting of the vector $Z'$ and the fermions $\chi$. The main
signature is $pp \to S$ production with an invisible decay to $Z' Z'
\to 4 \chi$.\bigskip

In essence, we find that the $Z'$ typically decays to SM fermions,
including a large branching ratio to leptons. If kinematically
possible, the invisible decay to dark matter will dominate, especially
for small mixing $\schi \lesssim 0.1$. In contrast, the new scalar
prefers decays to SM Higgs and gauge bosons, unless the decay $S \to
Z'Z'$ is kinematically allowed. The reason for this structure is that
all $Z'$ couplings with the exception to dark matter are mediated by
the mixing angle $\schi$. We will see that this structure inherently
limits discovery prospects for this kind of dark matter mediator at
colliders.

\subsection{$U(1)_{L_\mu-L_\tau}$}

\begin{figure}[t!]
\includegraphics[width=0.50\textwidth]{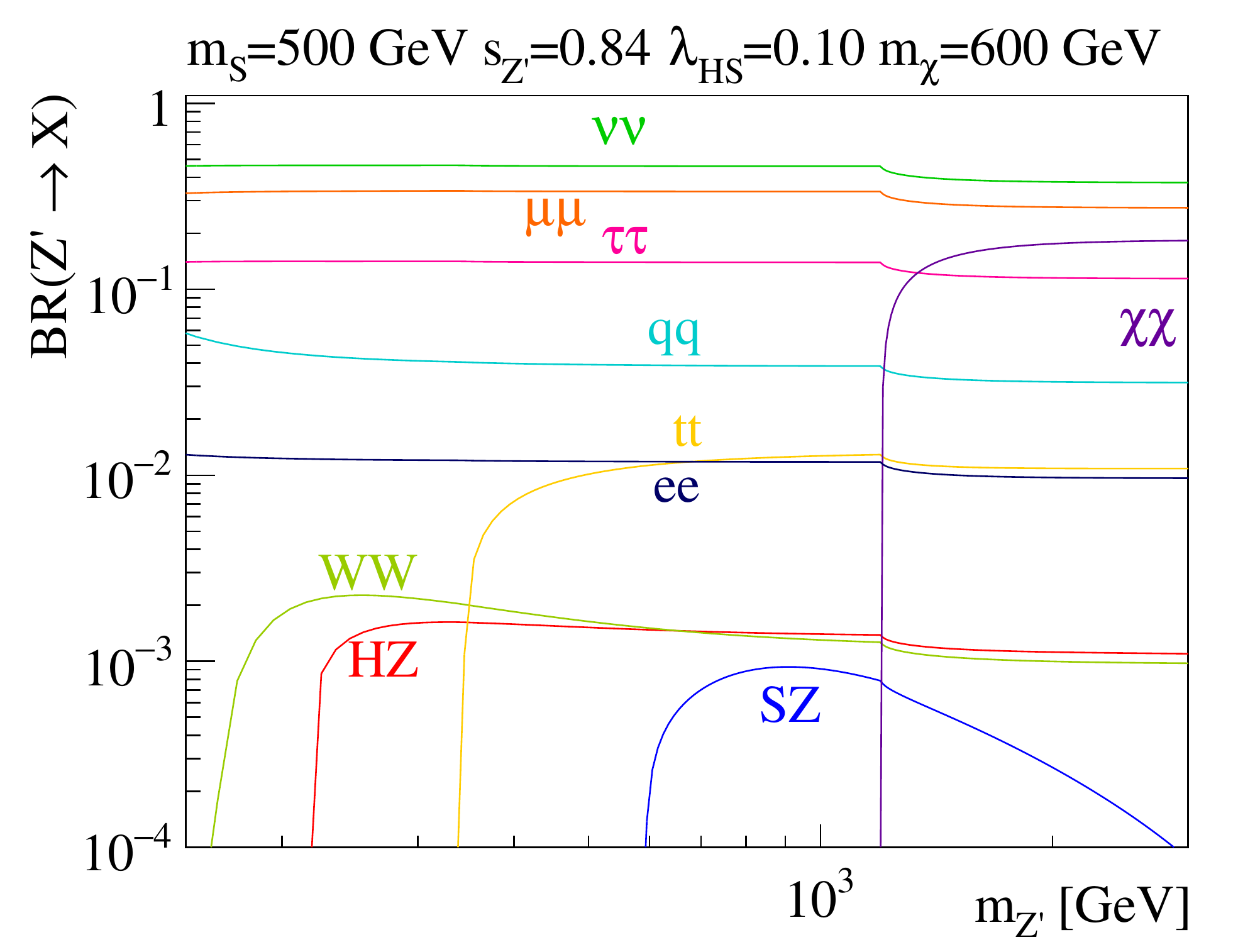}
\includegraphics[width=0.50\textwidth]{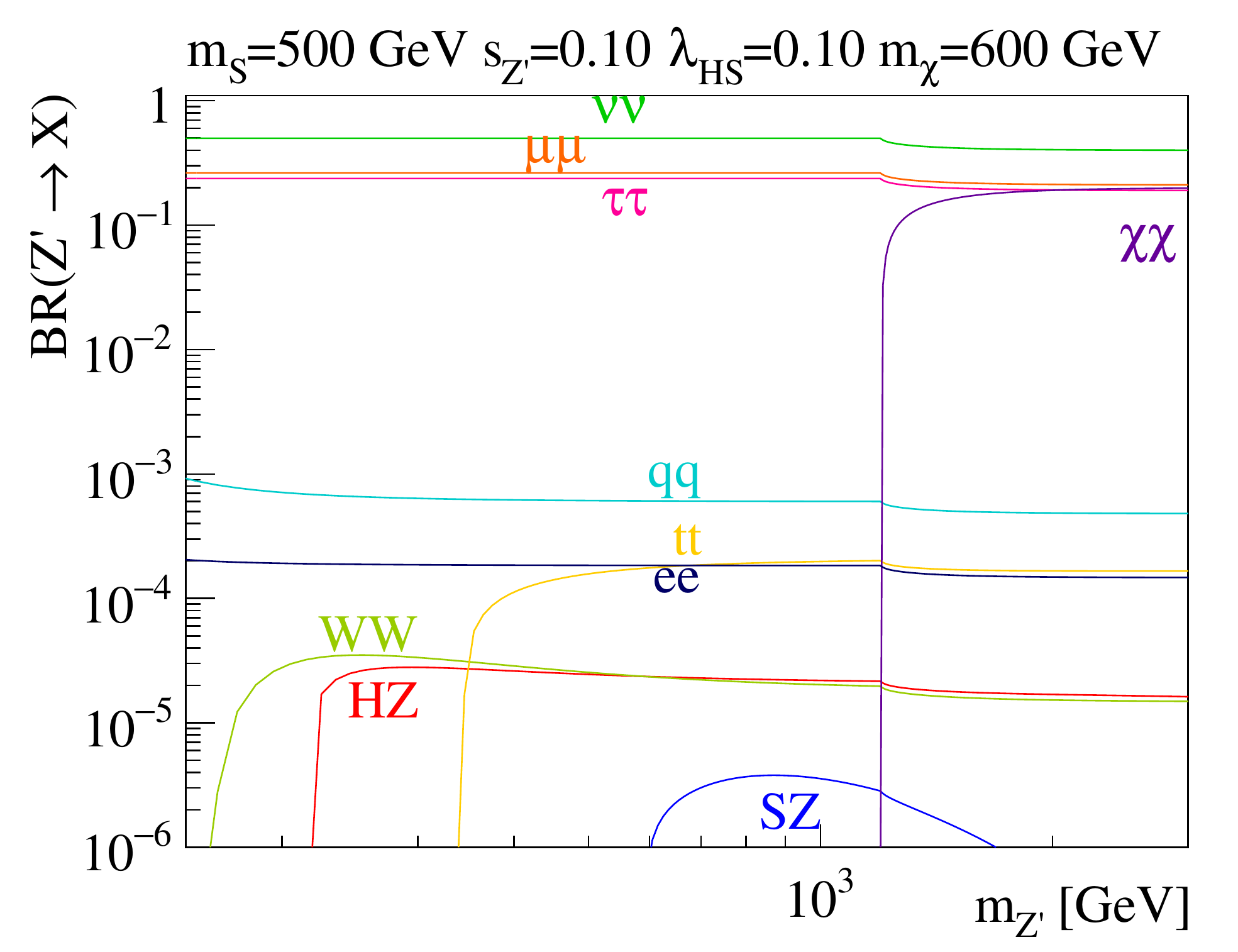}
\caption{Branching ratios for the $U(1)_{L_\mu-L_\tau}$ gauge boson
  with $m_S=500$~GeV, $\lambda_{HS}=0.1$, $g_{Z'}=1$, and $\schi =
  0.84$ (left) and $\schi = 0.1$ (right).}
\label{fig:ULmutauXBRs}
\end{figure}

Gauged differences of charged lepton numbers, such as
$U(1)_{L_\mu-L_\tau}$, induce $Z'$ gauge couplings to charged and
neutral leptons even for $\schi \to 0$. In return, SM lepton loops
generate kinetic mixing,
\begin{align}
\schi 
&= -\frac{3g' g_{Z'}}{4\pi^2} \int^1_0 dx \ x(x-1) 
   \log \frac{m_\tau^2+q^2x(x-1)}{m_\mu^2+q^2x(x-1)} \notag \\
&= \begin{cases}
\dfrac{g' g_{Z'}}{8\pi^2} \left( \dfrac{m_\tau^2}{q^2}-\dfrac{m_\mu^2}{q^2}\right)
  +\mathcal{O} \left(\dfrac{m_\tau^4}{q^4} \right)  & \quad \text{for } q^2\gg m_\tau^2\\[.5cm]
  \dfrac{g' g_{Z'}}{8 \pi^2} \log \dfrac{m_\tau^2}{m_\mu^2} + \mathcal{O}\left(\dfrac{q^2}{m_\mu^2}\right)
  \approx 0.025 \; g_{Z'}
 & \quad \text{for }q^2\ll m_\mu^2 \; .
  \end{cases}
\label{eq:kinloop}
\end{align}
Its size strongly depends on the energy scale at which we probe the
$Z$-$Z'$ mixing. At large momentum transfer, like at the LHC, the
mixing is dominated by the small ratio $m_\tau^2/q^2$. At low-energy
experiments, like direct dark matter detection, both leptons can be
integrated out and the remaining suppression is proportional to $\log
m_\tau^2/m_\mu^2$.  For an anomalous gauge group, this low-energy
limit would not be defined and instead require an additional physical
scale in the integral at which the anomaly is removed.

The fact that the loop-induced mixing is finite suggests that the
$U(1)_{L_\mu-L_\tau}$ gauge group can be embedded into a gauge group
which forms a direct product of $SU(3)_C\times SU(2)_L\times
SU(N)$~\cite{mu_tau_zprime_neut}. While in the unbroken phase of the
non-abelian group the additional condition $m_\mu = m_\tau$ removes
this contribution, it appears in the broken phase with
$U(1)_{L_\mu-L_\tau}$ intact.\bigskip

In the absence of kinetic mixing all couplings are fixed by the charge
assigned to the dark matter candidate. However, the LHC production
rate scales like
\begin{align}
\sigma(pp \to Z') \propto \schi^2 \; .
\end{align}
Hence, once the model predicts a sizeable LHC rate, searches for a di-lepton
resonance or for missing transverse energy are motivated by the
leading $Z'$ branching ratios. They are shown in
Fig.~\ref{fig:ULmutauXBRs}. For large mixing, the decays to muons and
taus and their neutrinos dominate, but branching ratios to di-jets
occur at per-cent level. Bosonic decays like $Z' \to HZ$ are rare, but
will be useful to disentangle the origin of the $U(1)$ structure. The
decay to two dark matter fermions opens above the kinematic threshold,
but remains below the neutrino contribution to the combined invisible
branching ratio. Reducing the mixing rapidly decouples all decay
signatures, with the exception of $Z' \to \mu \mu, \tau \tau, \nu \nu$
and $Z' \to \chi \bar{\chi}$.\bigskip

Altogether, similar to the $U(1)_X$ case we find that di-lepton
searches are the most promising ways to search for the $Z'$ boson. The
difference to the $U(1)_X$ case is the absence of lepton universality,
especially when it comes to electron couplings. In addition, the decay
to dark matter only dominates over the mixing-induced decay channels,
which also implies that the total invisible decay width tends to be
dominated by $Z' \to \nu \bar{\nu}$.

\subsection{$U(1)_{B-L}$}

\begin{figure}[t]
\includegraphics[width=0.48\textwidth]{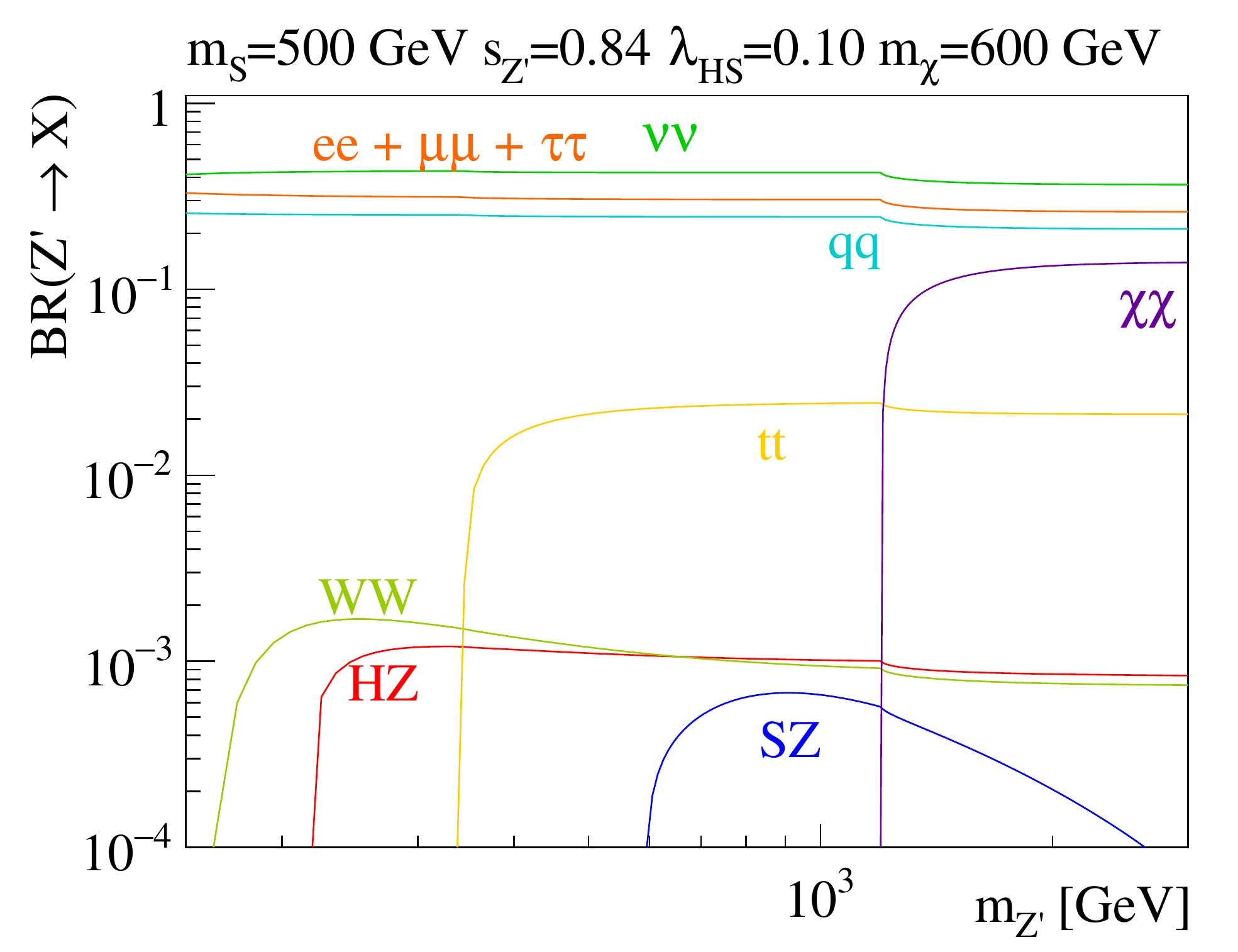}
\includegraphics[width=0.48\textwidth]{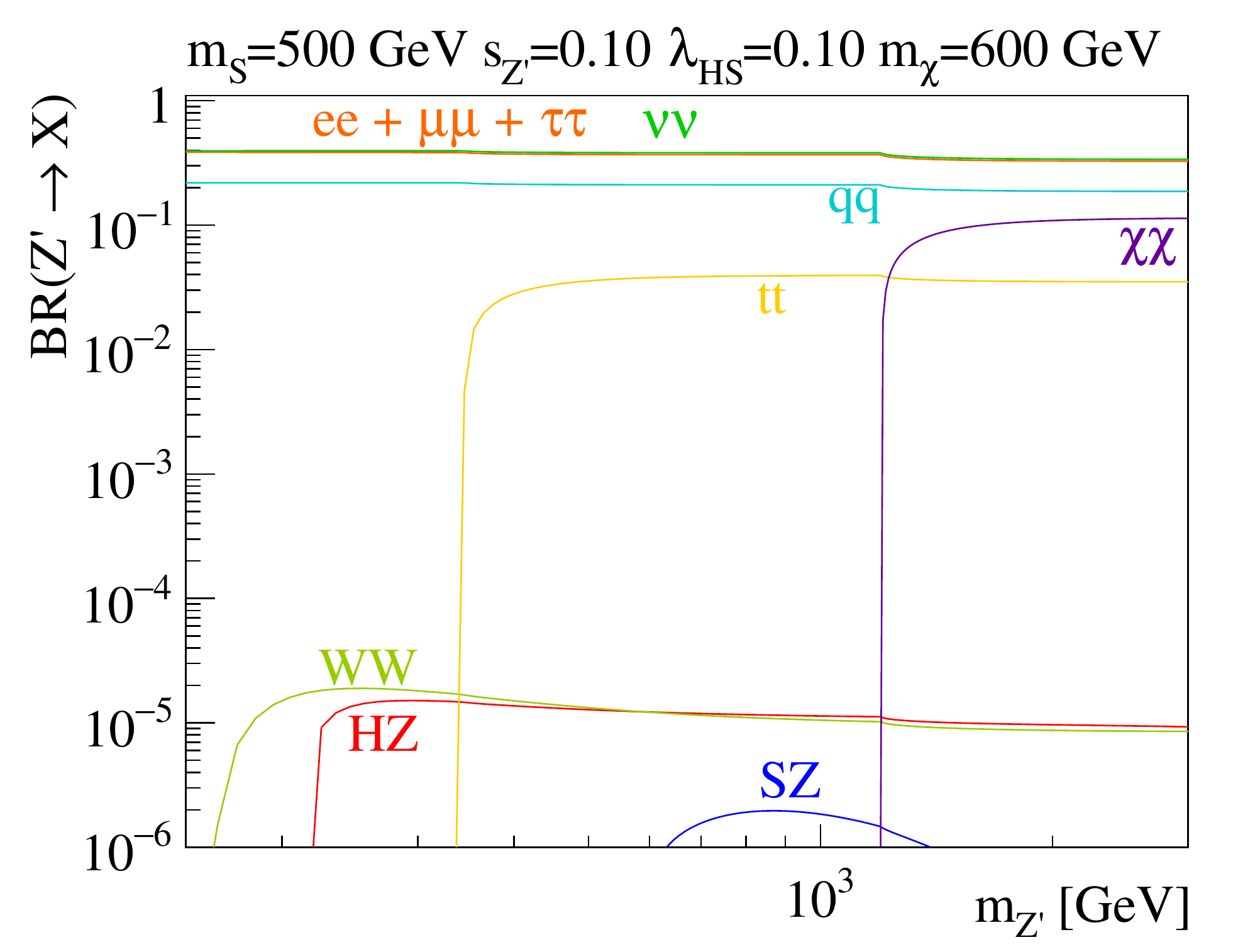}
\caption{Branching ratios for the $U(1)_{B-L}$ gauge boson with
  $m_S=500$~GeV, $\lambda_{HS}=0.1$, $g_{Z'}=1$, and $\schi = 0.84$
  (left) and $\schi = 0.1$ (right).}
\label{fig:UBminLBRs}
\end{figure}

The $U(1)_{B-L}$ gauge symmetry predicts new gauge couplings to both
quarks and leptons. Even for sizable kinetic mixing, the $Z'$
branching ratios shown in Fig.~\ref{fig:ULmutauXBRs} are largely
dictated by the charges,
\begin{align}
\br(Z'\to \ell^+\ell^-) : \br(Z'\to q\bar q) : \br(Z'\to \chi\bar \chi)
\approx n_\ell : \frac{n_q N_c}{9} : q_\chi^2 \; ,
\label{eq:bml_brs}
\end{align}
where $N_c$ is a color factor and the factor $1/9$ accounts for the
quark charges. We illustrate this scaling in
Fig.~\ref{fig:UBminLBRs}. Searches for di-lepton and di-jet resonances
are again promising. Even though a kinetic mixing of the kind shown in
Eq.\eqref{eq:kinloop} is induced, such a contribution hardly changes
the LHC search strategies, because tree-level generically beats
loops. The only difference between the two panels in
Fig.~\ref{fig:UBminLBRs} is that the bosonic channels decrease from
the per-mille level for large mixing to the $10^{-5}$ level for small
mixing. Invisible decays of the heavy vector boson will typically also
be dominated by $Z' \to \nu \nu$ decays, rather than decays to dark
matter, $Z' \to \chi \bar{\chi}$.\bigskip

From an LHC or relic density point of view the $U(1)_{B-L}$ scenario
is attractive, because the $Z'$ mediator has sizeable gauge couplings
to all fermions. For the phenomenology the universal mixing
contribution $\schi$ is generally sub-leading. The problem with this
model is that according to Fig.~\ref{fig:UBminLBRs} invisible $Z'$ decays
are dominated by decays to neutrinos. This means that an discovery of
a $Z'$ decaying invisibly might have nothing to do with dark
matter. 

\section{Collider and low-energy constraints}
\label{sec:constraints}

\begin{figure}[t!]
\includegraphics[width=1.00\textwidth]{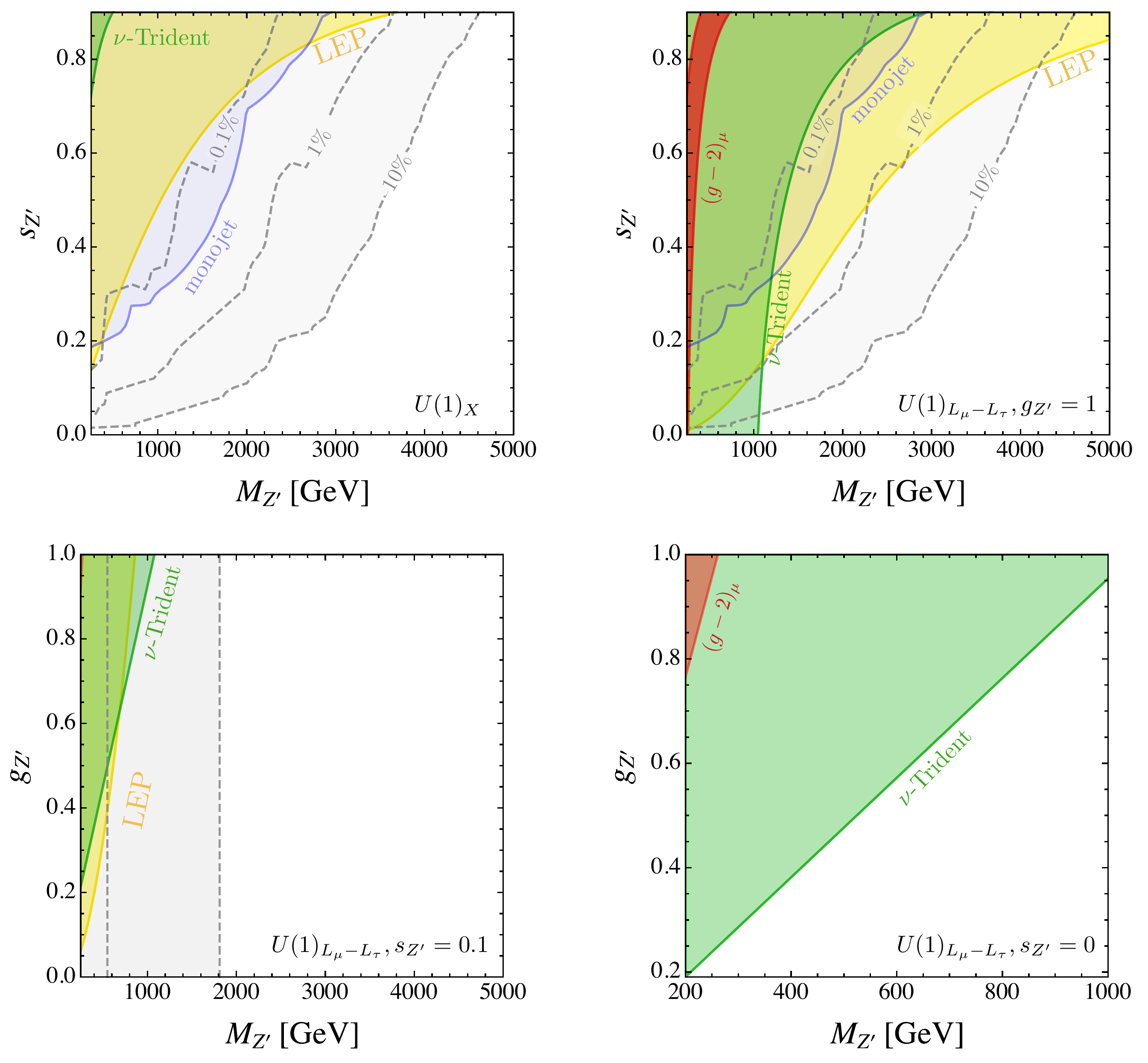}
\caption{Bounds on the kinetic mixing angle
  $\schi$, the mass $m_{Z'}$, and the gauge coupling $g_{Z'}$ for a
  $U(1)_X$ gauge boson (upper left) and a $U(1)_{L_\mu-L_\tau}$ gauge boson
  with $g_{Z'}=1$ (upper right) or $\schi=0.1$ (lower left) and $\schi=0$ (lower right). }\label{fig:limits1}
\end{figure}

New gauge bosons have motivated new physics searches for many
decades. For our three anomaly-free $U(1)$-extensions we consider
three different types of constraints: firstly, couplings to electrons
are constrained by LEP searches for new gauge bosons; secondly,
couplings to neutrinos lead to contributions to neutrino-nucleus
scattering; finally, couplings to light-flavor quarks predict sizeable
$Z'$ production rates at the LHC.

\subsection{LEP}
\label{sec:LEPconstraints}

Obviously, LEP strongly constrains the couplings of a new gauge boson for $m_{Z'}
\lesssim 209$~GeV. The luminosity at high energies
translates into a limit on the kinetic mixing angle mediating the
$Z'e^+ e^-$ interaction, namely $\schi < 0.03$.  This bound becomes
stronger for a lighter $Z'$~\cite{Hook:2010tw}.

Effects from heavier $Z'$ bosons can be described by effective 4-fermion
interactions~\cite{Han:2004az}
\begin{align}
\mathcal L_\text{eff} 
=&-\frac{\kappa_H}{2 \Lambda^2}  |H^\dagger D_\mu H|^2 \notag \\
&-\sum_{f,f'} \frac{4\pi\kappa_{ff'}}{\Lambda^2}(\bar f\gamma^\mu f)(\bar f'\gamma_\mu f')
-\sum_f \left( \frac{i\kappa_{Hf}}{\Lambda^2} (\bar f\gamma^\mu f)(H^\dagger D_\mu H) 
               + \text{h.c.} \right) \; .
\end{align}
Any $Z'$ couplings involving the Higgs are either proportional to the
scalar mixing angle $s_\alpha$ or of higher order in $v/v_S$ or
$\schi$.  As for searches for contact interactions, the strongest
constraints arise from $e^+e^- \to q \bar q, \ell^+ \ell^-$
searches~\cite{Alcaraz:2006mx, Schael:2013ita}. At $95\%$ C.L. the LEP
limits are
\begin{alignat}{9}
\frac{\Lambda}{\sqrt{\kappa_{\ell\ell}}}  &\gtrsim 24.5 \;\tev 
&\qqqquad 
\frac{\Lambda}{\sqrt{\kappa_{e\mu}}}  &\gtrsim 18.6 \;\tev \notag \\
\frac{\Lambda}{\sqrt{\kappa_{e\tau}}}  &\gtrsim 15.6 \;\tev
&\qqqquad 
\frac{\Lambda}{\sqrt{\kappa_{eu}}}  &\gtrsim 14 \;\tev \; .
\label{eq:lep_fourfermion}
\end{alignat}
Matching to the full theory we identify the new physics scale as
$\Lambda = \sqrt{8\pi} m_{Z'}$ and the Wilson coefficients
$\kappa_{ff'}$ as functions of the $Z'$ couplings, the mixing angle
$s_{Z'}$ and the chirality of the involved fermions. These LEP
constraints put strong bounds on the new gauge bosons couplings to
electrons~\cite{BLlimits},
\begin{align}
\frac{m_{Z'}}{g_{Z'}}&> 6.9 \;\tev \qqquad &&U(1)_{B-L}  \notag \\
\frac{m_{Z'}}{g_{Z'}}&> 5.25\;\tev \qqquad &&U_{L_e-L_\mu}, U_{L_e-L_\tau} \; .
\label{eq:etaub}
\end{align}
These limits for $s_{Z'}=0$ become even stronger for $s_{Z'}>0$.  The
remaining parameter space will typically not give the observed relic
density and push the additional scalar $S$ to large masses. This is
why at this stage we will drop the $U(1)_{B-L}$ gauge group (and any
other group with gauged electrons) from our analysis.\bigskip

For the remaining gauge groups $U(1)_X$ and $U_{L_\mu-L_\tau}$ the
coupling to leptons and with it the sensitivity to LEP constraints
depends on the mixing angle.  In Fig.~\ref{fig:limits1} we show the
excluded parameter space for $U(1)_X$ and $U_{L_\mu-L_\tau}$, the
latter for fixed $g_{Z'}=1$ or $\schi=0.1$. First, we see that the
the four-fermion constraints constrain both, the $U(1)_X$ and the
$U_{L_\mu-L_\tau}$ models, unless $s_Z'=0$. In terms of $m_{Z'}$ and $\schi$ the limits
on both gauge groups are similar, because in both cases the electron
couplings enters with $\schi$, but the muon coupling is a gauge
coupling for $U_{L_\mu-L_\tau}$.\bigskip

In the $U(1)_X$ model, the LEP constraints from contact interactions
on $m_{Z'}$ and $s_{Z'}$ are similar in strength to the bound from the
modification of the $Z$ mass,
Eq.\eqref{eq:zmassconstraint}~\cite{ew_precision}, but stronger for
the $U(1)_{L_\mu-L_\tau}$ gauge boson with a coupling $g_{Z'}\gtrsim
0.5$. Additional bounds arise from non-universal $Z$-couplings to
muons and electrons
This constraint is weaker than both the bounds from four-fermion
interactions and from the $Z$-mass measurement, but in
$U(1)_{L_\mu-L_\tau}$ a contribution arises at the one-loop level from $Z'$ exchange between the muon legs, which is present in the limit $s_{Z'}\to
0 $ as well. The corresponding constraints are however weaker than the dominant constraint from
neutrino-trident production discussed in the following
section~\cite{trident}.

\subsection{Low-energy probes}
\label{sec:lowenergyconstraints}

\begin{figure}[t]
\centering 
\begin{fmfgraph*}(100,80)
\fmfstraight
\fmfset{arrow_len}{2mm}
\fmfleft{i1,i2}  
\fmfright{o1,o2,o3,o4}
\fmf{fermion,width=0.6,tension=1}{i1,vn}
\fmf{fermion,width=0.6}{vn,o1}
\fmffreeze
\fmf{boson,width=0.6,lab.side=left,label=$\gamma$}{vn,vm1}
\fmf{fermion,width=0.6}{vm1,o2}
\fmf{fermion,width=0.6}{vm2,vm1}
\fmf{fermion,width=0.6}{vm2,o3}
\fmf{fermion,width=0.6,tension=2}{i2,vl}
\fmf{fermion,width=0.6}{vl,o4}
\fmf{boson,width=0.6,lab.side=left,label=$Z'$}{vm2,vl}
\fmflabel{$N$}{i1}
\fmflabel{$\nu$}{i2}
\fmflabel{$N$}{o1}
\fmflabel{$\mu^-$}{o2}
\fmflabel{$\mu^+$}{o3}
\fmflabel{$\nu$}{o4}
\end{fmfgraph*}
\caption{Example Feynman diagrams for neutrino trident production.}
\label{fig:feyn_trident}
\end{figure}
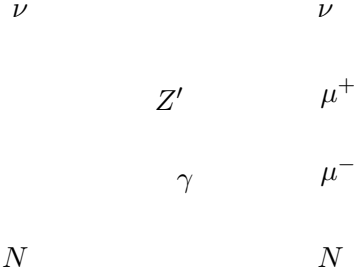

Additional gauge bosons are constrained by wealth
of low-energy experiments. In our case, the $U(1)_{L_\tau-L_\mu}$ and
$U(1)_{B-L}$ gauge bosons contribute to the production of $\mu^-\mu^+$
pairs in neutrino--nucleus scattering or neutrino trident production
\begin{align}
\nu_\mu N\rightarrow \nu_\mu N \; \mu^+\mu^- \; , 
\end{align}
shown in Fig.~\ref{fig:feyn_trident}.  The enhancement over the SM
prediction for the total trident cross section to the SM prediction in
the limit $m_{Z'} \gg m_\mu$,
is~\cite{trident}
\begin{align}
\frac{\sigma}{\sigma_\text{SM}} =
\frac{(1+2\, C_A^{Z'})^2+(1+4s_w^2+2\, C_V^{Z'})^2}{1+(1+4s_w^2)^2}\,,
\end{align}
with
\begin{align}\label{eq:tridentcoeff}
C_V^{Z'}&=\frac{v^2}{4c_{Z'}^2m_{Z'}^2}\bigg(4g_{Z'}^2+5g'g_{Z'}s_{Z'}+\frac{3}{2}g^{\prime \, 2}s_{Z'}^2\bigg)\,,\\
C_A^{Z'}&=\frac{v^2}{4c_{Z'}^2m_{Z'}^2}\bigg(g'g_{Z'}s_{Z'}+\frac{1}{2}g^{\prime\, 2} s_{Z'}^2\bigg)\,.
\end{align}
In our evaluation we neglect
corrections $v/v_S$, and for the $U(1)_X$ gauge group all terms
proportional to the new gauge coupling $g_{Z'}$ vanish. 
The combined measurement from CHARM-II~\cite{Geiregat:1990gz} and
CCFR~\cite{Mishra:1991bv} comes to
\begin{align}
\frac{\sigma}{\sigma_\text{SM}} = 0.83 \pm 0.28 \; .
\end{align}
We show the excluded parameter space in Fig.~\ref{fig:limits1}. For
$U(1)_X$ the trident constraints are mediated by the
two mixing vertices, so the cross section is suppressed by $\schi^4
\ll 1$. For $U(1)_{L_\mu-L_\tau}$ all vertices are new gauge
couplings, so the trident constraint becomes much stronger and survives 
the limit $\schi \to 0$.\bigskip

Interestingly, the $U(1)_{{L_\mu}-L_{\tau}}$ gauge boson can also provide
an explanation of the long-standing discrepancy between the
experimental value and the SM prediction for the anomalous magnetic
moment of the muon~\cite{g-2}
\begin{align}
a_\mu^\text{exp}-a^\text{SM}_\mu = (29.3 \pm 7.6) \times 10^{-10} \; . 
\end{align}
The $Z'$ contribution in the limit $m_{Z'} \gg m_\mu$ is given
by~\cite{Baek:2001kca}
\begin{align}
\Delta a_\mu = \frac{1}{48\pi^2}\frac{m_\mu^2}{m_{Z'}^2}\frac{1}{c_{Z'}^2}\bigg({g'}^2s_{Z'}^2+6g'g_{Z'}\schi+4 g_{Z'}^2\bigg)\,.
\end{align}
We show the preferred region shaded in red in Fig.~\ref{fig:limits1}. In
all cases, this explanation is excluded for the masses we consider.
Other low-energy constraints such as lepton flavor universality in
$\tau$ decays or atomic parity violation do not yield additional
constraints for the models and the parameter spaces we consider.

\subsection{LHC resonance searches}
\label{sec:lhcconstraints}

Especially for heavier resonances, LHC searches for di-jet and
di-lepton resonances constraint the mass range
$m_{Z'}=250~...~5000$~GeV~\cite{ATLAS:2016cyf, Sirunyan:2016iap},
provided there is a large enough coupling to the incoming quarks. In
the case of the $U(1)_{B-L}$ gauge boson, the production cross section
and all decay channels are sensitive to the universal coupling
$g_{Z'}$~\cite{Okada:2016gsh}, unless $g_{Z'}\ll 1$ makes it hard to
obtain the correct dark matter abundance.  For $U(1)_X$ or
$U(1)_{L_\mu-L_\tau}$ gauge bosons the production cross section at the LHC depends on the kinetic mixing angle,
\begin{align}
\sigma( q\bar{q} \to Z')=
\frac{\pi^2}{12 s} \; \frac{\alpha_e}{c_w^2} \; \tchi^2 \;
\sum_q \left( Q_q^2+(T_3-Q_q)^2 \right) \; ,
\end{align}
where $Q_q$ and $T_3$ are the electric charges and the weak isospin of
the quarks we neglect corrections of order $v^2/v_S^2$.

In Fig.~\ref{fig:limits1} we include some approximate LHC limits for
illustration.  We compute the $Z'$ production cross section with
\textsc{MadGraph5}~\cite{madgraph}, accounting for higher order
corrections using \textsc{Matrix}~\cite{Hamberg:1990np,
  Grazzini:2017mhc}, estimating the NNLO effects by using the K-factor for the $Z$ boson Drell-Yan production cross section, and compare with the ATLAS di-lepton
limits~\cite{ATLAS:2016cyf}. We take the branching ratio
$\br(Z'\to \ell^+\ell^-)$ to be a free parameter and we show the
excluded parameter space for values of 0.01\% and 1\% for $U(1)_X$ and
1\% and 10\% for $U(1)_{L_\mu-L_\tau}$. Especially in the
$U(1)_{L_\mu-L_\tau}$ case a strong suppression of the decay $Z'\to
\mu^+\mu^-$ rate can only be achieved through a large
$U(1)_{L_\mu-L_\tau}$ charge of the dark matter candidate, leading to
a Landau pole of the $U(1)_{L_\mu-L_\tau}$ gauge couplings at low
energies.
 
In Fig.~\ref{fig:limits1} we see that for small mixing angles all LHC
constraints vanish, because the quarks are neutral under the two gauge
groups. In the right panel we show that for a fixed, but small mixing
angle the LHC production rate is fixed as well, and for a fixed
branching ratio to leptons the allowed $Z'$ masses are typically in
the TeV-range. Lighter new gauge bosons are only allowed for small
mixing angles $\schi < 0.1$.

LHC searches for invisible $Z'$ decays will be discussed in
Sec.~\ref{sec:lhc_inv}.  Searches for $Z\to 4\mu$ decays at the LHC
can lead to additional constraints in the case of the
$U(1)_{L_\mu-L_\tau}$ gauge boson for $\schi \ll 1$, but the
corresponding parameter space is excluded by the neutrino trident
constraint discussed above for the masses we consider \cite{CMS:2018poc}.

\section{Dark matter constraints}
\label{sec:dm}

If we consider our $Z'$ models to be consistent and realistic, they have to
reproduce the observed relic density, or at least predict a
sufficiently large annihilation rate after thermal decoupling. We will
see that explaining only a fraction of the observed dark matter does
not circumvent the constraints, because the typical problem is to
reach a large enough dark matter annihilation rate. Given that our
model is meant to explain the observed relic density, it then has to
respect constraints from indirect and direct detection experiments.

\begin{figure}
\includegraphics[width=1.0\textwidth]{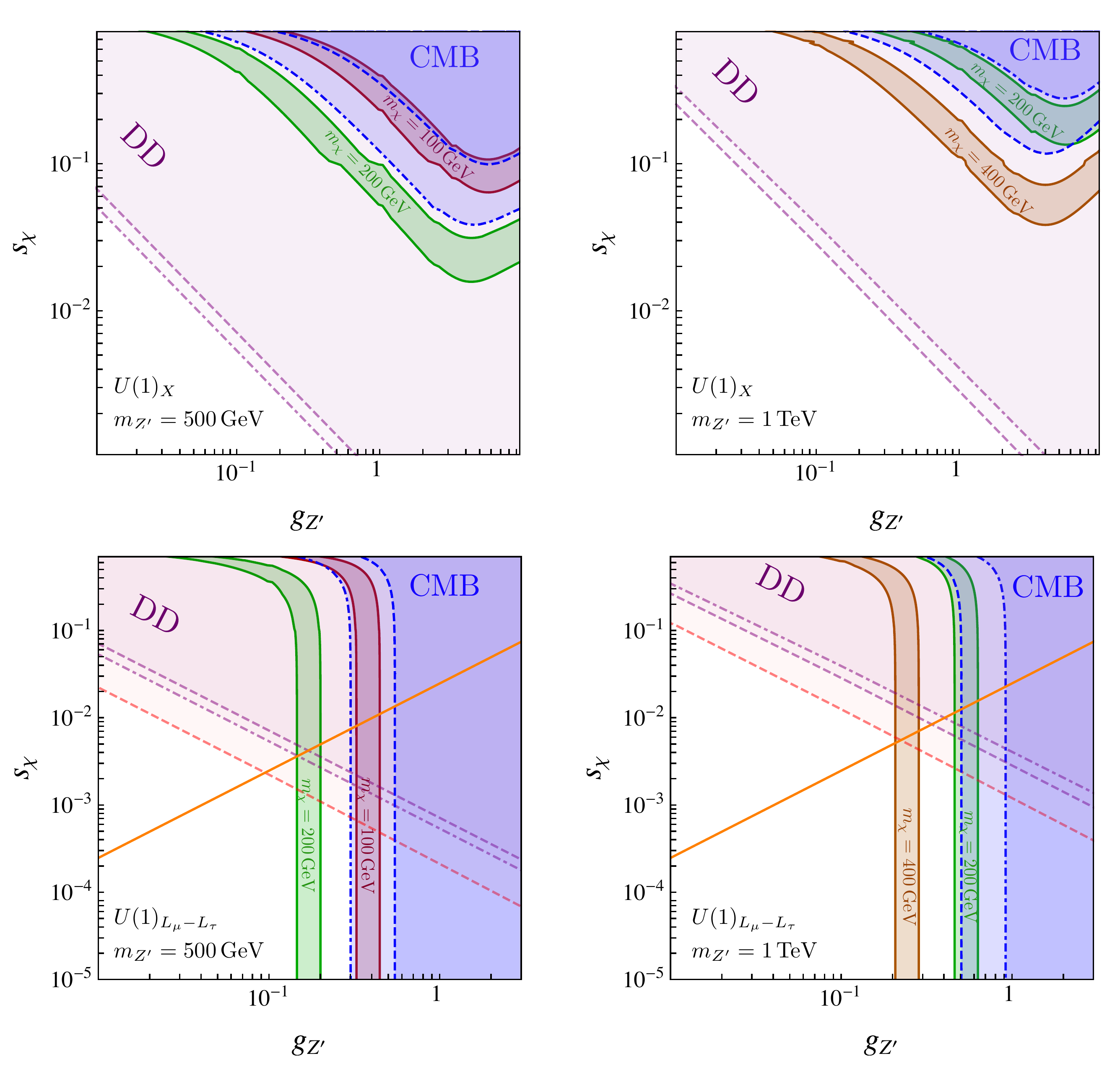}
\caption{Constraints from the observed relic density in the $U(1)_X$
  (upper) and $U(1)_{L_\mu-L_\tau}$ model (lower). We fix
  $m_{Z'}=500$~GeV (left) with $m_\chi=100, 200$~GeV or $m_{Z'}=1$~TeV
  (right) with $m_\chi=200, 400$~GeV. Indirect and direct detection
  constraints are shaded. The (dot-) dashed contours correspond to
  $m_\chi=(200) 100$~GeV or $m_\chi=(400) 200$~GeV, respectively. The
  Xenon1T projection is indicated by the red dashed contour in the
  lower panels. The orange line indicates the purely loop-induced
  mixing angle for a given $g_{Z'}$.}
\label{fig:relic1}
\end{figure}

\subsection{Relic density}
\label{sec:dm_relic}

Dark matter annihilation is dominated by $Z'$ in the $s$-channel,
\begin{align}
\chi \bar \chi \to Z' \to \text{SM} \; .
\end{align}
The scalar $S$ has no direct couplings to dark matter, so there is no $S$
mediated annihilation and the scalar only plays a role
in the annihilation channel
\begin{align}
\chi \bar \chi\to Z' \to Z S \; .
\end{align}
For the $U(1)_X$ model, the gauge coupling $g_{Z'}$ and the mixing
angle $\schi$ both enter the annihilation rate, because they determine
the $Z'$ coupling to dark matter and SM particles, respectively. In
the upper panels of Fig.~\ref{fig:relic1} we show the corresponding
parameter space, for which a relic density in the range
$(0.3~...~1.1)\times 0.12$~\cite{planck} is reproduced for
$m_{Z'}=500$~GeV, with $m_\chi=100$~GeV or $m_\chi=200$~GeV (upper left) and $m_{Z'}=1$~TeV, with $m_\chi=200$~GeV or $m_\chi=400$~GeV (upper right), respectively. Apart
from near the $Z'$-pole $m_{Z'} = 2 m_\chi$, both couplings need to be
sizable to reproduce the observed relic abundance.  Generally, a large
gauge coupling $g_{Z'}$ allows for smaller mixing angles $\schi$; only
for very large $g_{Z'}$ the mixing angle $\schi$ has to increase again
to introduce a sizable $Z'$ branching ratio to SM particles.
Following Eq.\eqref{eq:mass_ratio} this constrains the mass splitting
between the $Z'$ and the scalar $S$ mediators. If we assume $m_\chi =
200$~GeV we find that $g_{Z'} = 0.1~...~1$ requires roughly $s_{Z'} =
0.5~...~0.04$, translating into
%
\begin{align}
\frac{m_S}{m_{Z'}} = ( 1~...~6 ) \; \sqrt{\lambda_S} \; ,
\end{align}
if, following Eq.\eqref{eq:qs_one}, we assume $q_S = 1$ based on the
neutrino sector.

In the $U(1)_{L_\mu-L_\tau}$ case, the relic density can be set by
$g_{Z'}$ alone, as is evident from the lower panels of
Fig.~\ref{fig:relic1}. In the absence of additional matter, the mixing
angle for a given gauge coupling $g_{Z'}$ is specified by
Eq.\eqref{eq:kinloop}. We indicate this loop-induced value of the
mixing angles $s_{Z'}$ for gauge couplings preferred by the relic
density by the orange line in the lower panels of
Fig.~\ref{fig:relic1}.  For the mass splitting between the two
mediators we now find
\begin{align}
\frac{m_S}{m_{Z'}} \approx \frac{\sqrt{\lambda_S}}{g_{Z'}}
                   = ( 0.2~...~0.5 ) \; \sqrt{\lambda_S} \; .
\end{align}
\bigskip

A general bound on the mass of the dark matter candidate arises from
the bound on invisible Higgs decays $\br(H \rightarrow
\text{inv})<0.23$~\cite{hinv,legacy}. It constrains the loop-induced
decay $H\to \chi \bar \chi$ through the $H-Z'-Z'$ coupling. We avoid
this constraint by assuming $2m_\chi > m_H$. 

\subsection{Indirect detection}
\label{sec:dm_id}

If dark matter annihilates into charged leptons, it can be constrained
through the cosmic positron flux. The positron spectrum has been
measured by HEAT~\cite{DuVernois:2001bb},
PAMELA~\cite{Adriani:2013uda}, FERMI-LAT~\cite{FermiLAT:2011ab}, and
AMS~\cite{Accardo:2014lma}. It is most sensitive to dark matter masses
around 100~GeV. For heavier dark matter the sensitivity drops
rapidly~\cite{ams_theo}, and uncertainties in the astrophysical
background modeling translate into sizable errors in the production
cross section and slope of the measured
spectrum~\cite{Delahaye:2008ua}. Note that we do not attempt a fit of
excesses in PAMELA, FERMI-LAT or AMS~\cite{Baek:2008nz}.\bigskip

An especially clean test of many dark matter models is provided by
measurements of the polarization fluctuation and temperature of the
cosmic microwave background (CMB)~\cite{Slatyer:2009yq}. Dark matter
annihilation during the period of last scattering induces distortions
of the CMB spectrum and temperature. Annihilation into charged
leptons, in particular electrons, comes with the highest effective
deposited power fraction $f_\text{eff}$. A dominant
annihilation channel driven by large kinetic mixing in the $U(1)_X$ and
$U(1)_{L_\mu-L_\tau}$ models is 
\begin{align}
\chi \bar \chi \to Z'\to e^+ e^- \; ,
\end{align}
driven by 
the significant coupling of the $Z'$ to the electromagnetic current.
For 
$U(1)_{L_\mu-L_\tau}$ the limit $\schi=0$ leaves us with annihilation 
into muons, taus, and neutrinos. The current
limit obtained from Planck data on the annihilation cross section
reads~\cite{planck, Altmannshofer:2016jzy}
\begin{align}
f_\text{eff} \; \frac{\sigma v}{m_\chi}\lesssim 3 \times 10^{-28}\,\frac{\text{cm}^3}{\gev s}\,.
\end{align}
A conservative bound assumes $100\%$ annihilation into electrons,
unless $\schi<0.1$. For $U(1)_{L_\mu-L_\tau}$, we assume a dominant
annihilation into muons. The corresponding limits are shown in
Fig.~\ref{fig:relic1} shaded blue with dashed and dot-dashed contours
for $m_\chi =100$~GeV and $m_\chi =200$~GeV ($m_\chi =200$~GeV and
$m_\chi =400$~GeV), respectively.

\subsection{Direct detection}
\label{sec:dm_dd}

Direct detection experiments are sensitive to dark matter scattering
off heavy nuclei through $Z'$ exchange, specifically spin-independent
scattering in analogy to Higgs exchange.  The strongest bounds on
spin-independent scattering come from LUX~\cite{Akerib:2016vxi},
PANDA-X II~\cite{Tan:2016zwf} and Xenon1T~\cite{Aprile:2017iyp}. 

In Fig.~\ref{fig:relic1}, we show the constraints obtained by the
first Xenon1T results for the $U(1)_{X}$ extension (upper panels)
and the $U(1)_{L_\mu - L_\tau}$ extension (lower panels).  
%
%
The
excluded region is indicated in purple, with dashed and dot-dashed
contours for $m_\chi =100 (200)$~GeV and $m_\chi =200 (400)$~GeV,
respectively. We further include the projected reach for
XenonnT~\cite{xenonnt} in the lower panels for $m_\chi =200 (400)$~GeV
as a dashed red contour.  

For both models, the $Z'$ couplings to nuclei are proportional to the
kinetic mixing $\schi$. In the $U(1)_X$ model the values of $\schi$
necessary to explain the relic density are completely excluded by
Xenon1T. In contrast, for the $U(1)_{L_\mu-L_\tau}$ model the relic
density can be set by annihilation through the gauge coupling $g_{Z'}$
alone, while the direct detection cross section is proportional to
$s_{Z'}$.  In absence of a tree-level mixing, the loop-induced mixing
given in Eq.\eqref{eq:kinloop} is the largest effect from
$g_{Z'}$-dependent couplings. Couplings not proportional to the
kinetic mixing only arise at the two-loop level~\cite{Kopp:2009et} and
can be neglected. We indicate the value of the loop-induced mixing
angles in Fig.~\ref{fig:relic1} as an orange line. For both
$m_{Z'}=500$ GeV and $m_{Z'}=1$ TeV, a purely loop-induced kinetic
mixing allows for an explanation of the observed DM relic density.

\section{LHC signatures}
\label{sec:lhc}

A key question for $Z'$ mediators at the LHC is how we can establish
the link to the dark matter sector once we discover a di-lepton
resonance through kinetic mixing.  This is complicated by the presence
of sizable $Z'$ branching ratios to neutrinos in the $U(1)_X$ and
$U(1)_{L_\mu-L_\tau}$ models. We follow two strategies to establish
the $Z'$ as a dark matter mediator: a profile analysis of the
di-lepton mass peak~\cite{ALEPH:2005ab} and a combination with the
mono-jet signal. In the case of very small mixing angles the production cross section of the $Z'$ can become smaller than the production cross section of the scalar $S$, whose decays are dominated by the $S\to Z'Z'$ decay rate. We present a third discovery strategy based on the process $S\to Z'Z'\to \mu^+\mu^- \met$.\bigskip

For any thermal dark matter scenario, the relic abundance strongly
constrains the kinetic mixing angle $s_{Z'}$.  As discussed in the
last section, a $U(1)_X$ gauge boson is excluded as a single mediator
through direct detection.  For a $U(1)_{L_\mu-L_\tau}$ gauge boson
with $m_{Z'}\lesssim 1$~TeV direct detection requires $s_{Z'}\lesssim
0.01$, leading to a suppressed $Z'$ production rate. In addition, the
gauge coupling needs to be sizable $g_{Z'}>0.1$, to allow for an
efficient annihilation in the early universe.  Following
Eq.\eqref{eq:mass_ratio} the scalar $S$ then cannot decouple from the
spectrum and will therefore play an important role in the LHC
phenomenology.

\subsection{$Z'$ profile}
\label{sec:lhc_profile}

In both, the $U(1)_X$ and the $U(1)_{L_\mu-L_\tau}$ models, the
mediator has a sizable branching ratio into leptons. We can
approximately relate the di-lepton production rate to the mono-jet
signal via
\begin{align}
\frac{\sigma(pp \to Z'\to \met+\text{jet})}
     {\sigma(pp \to Z' \to \ell^+\ell^-)}
=\frac{\alpha_s}{4\pi}
 \frac{\br(Z'\to \chi\bar \chi)+\br(Z' \to \nu\bar \nu)}
 {\br(Z' \to \ell^+\ell^-)}\; .
\label{eq:lhc_signals}
\end{align}
It is safe to assume that any kinetic mixing large enough to observe a
mono-jet signal will first give a di-lepton signal.

\begin{figure}
\centering
\includegraphics[width=\textwidth]{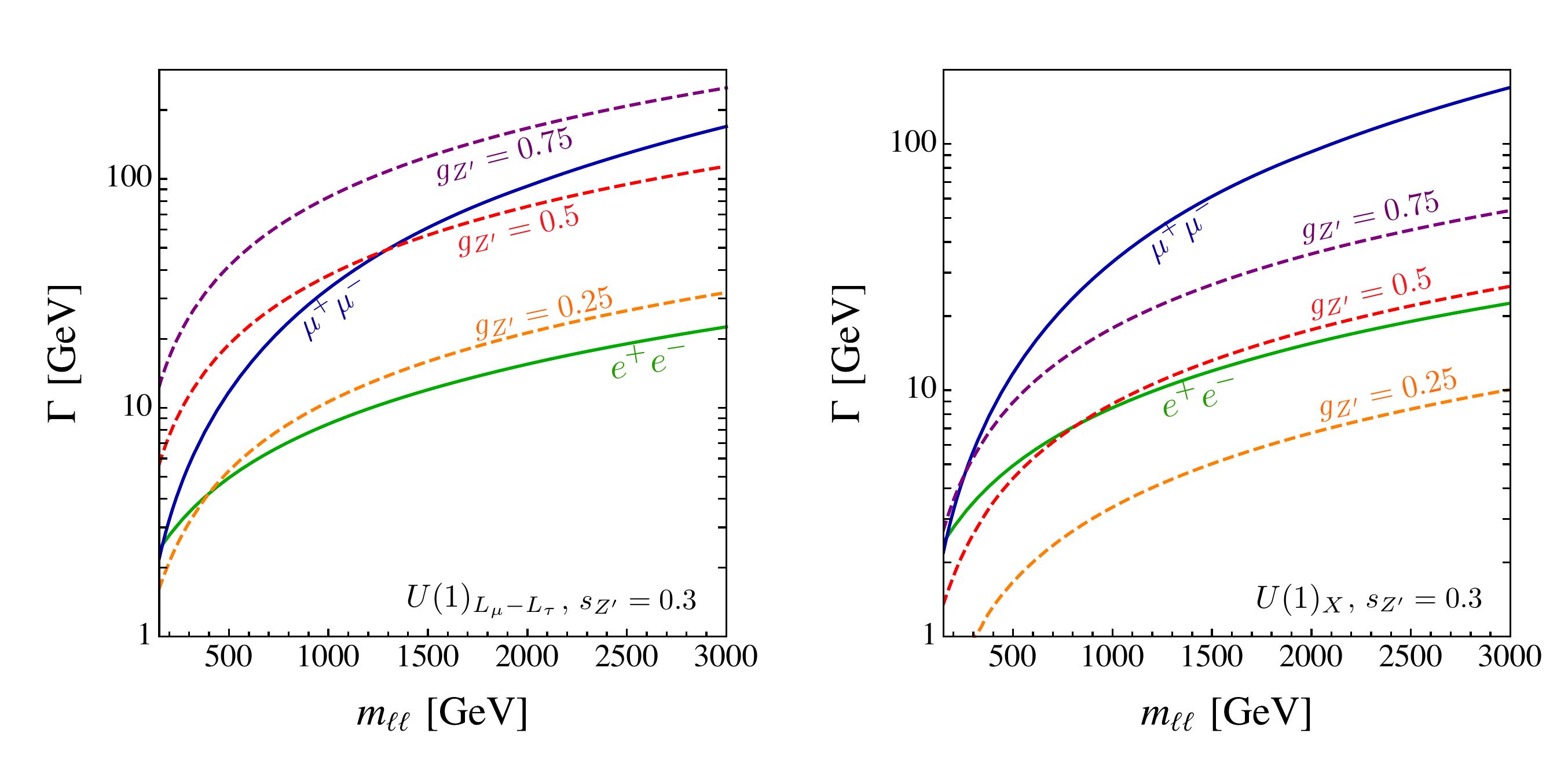}
\caption{Total $Z'$ width predicted in the $U(1)_X$ and
  $U(1)_{L_{\mu}-L_\tau}$ models for $s_{Z'}=0.3$ and $g_{Z'}=0.25,
  0.5, 0.75$. We also show the detector resolution for $e^+e^-$ and
  $\mu^+\mu^-$ resonances as a function of the di-lepton mass
  identifying $m_{Z'} = m_{\ell\ell}$.}
\label{fig:widths}
\end{figure}

In this situation, we can use a fit of the $Z'$-width in the di-lepton
channel to constrain the $Z'$ branching ratio to dark matter, in
analogy to the measurement of the number of light neutrinos at
LEP \cite{ALEPH:2005ab}. This measurement heavily relies on the ATLAS and CMS energy
resolution for high-energy di-leptons.  The lepton energy resolution
translates into a resolution of the $Z'$ width at the per-cent level
for electrons~\cite{hayden_thesis} and several per-cent for
muons~\cite{Radogna:2016jsf}.  In Fig.~\ref{fig:widths} we compare the
experimental resolution for $Z'\to \mu^+\mu^-$ and $Z'\to e^+e^-$ to
the predicted $Z'$ width in the $U(1)_{L_\mu-L_\tau}$ model (left) and
$U(1)_X$ model (right) for $g_{Z'}=0.25, 0.5, 0.75$. A shape analysis
will only give information on invisible $Z'$ decays if detector
resolution is smaller than the total width.
 
In the $U(1)_{L_\mu-L_\tau}$ model, the branching ratio $\br(Z'\to
e^+e^-)$ is suppressed by the kinetic mixing $s_{Z'}$. A fit to the
$Z'$ width in this channel can still constrain an invisible $Z'$ decay
channel to 1\% or better.

\subsection{Invisible $Z'$ decays}
\label{sec:lhc_inv}

\begin{figure}[t]
\begin{center}
\includegraphics[width=1.05\textwidth]{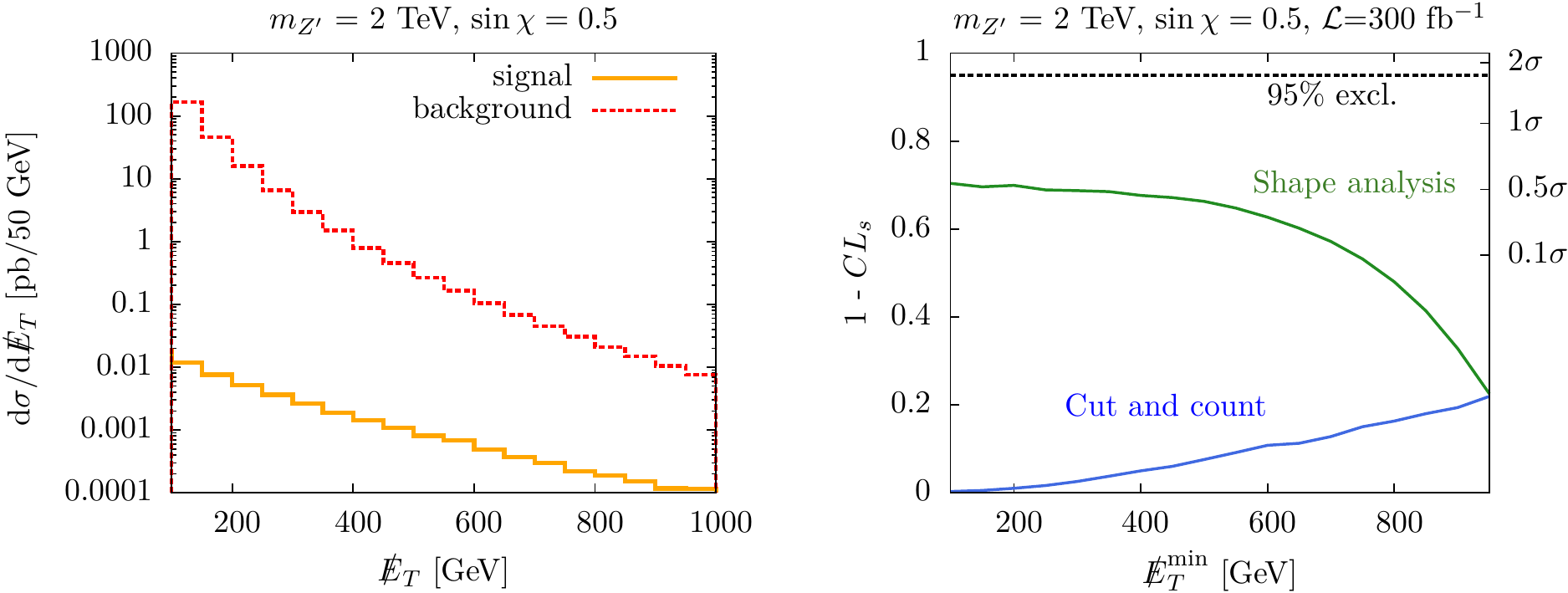}
\end{center}
\caption{Left: $\met$ distribution for a typical signal and the
  combined $W/Z$+jets background. Right: expected confidence limit and
  associated Gaussian significance as a function of the $\met$ cut
  using the full $\met$ shape information (green) versus a
  cut-and-count results (blue).}
\label{fig:etmiss} 
\end{figure}

An alternative strategy to establish the nature of the $Z'$ as a dark
matter mediator is to measure the mono-jet cross section and combine
it with the di-lepton rate. The presence of a dark matter coupling
strongly enhances the predicted invisible $Z^\prime$ width. For
instance, for the $U(1)_X$ model typically $\br (Z'\to \nu \bar
\nu)\approx 10\%$ without any coupling to dark matter and $\br (Z'\to
\chi \bar \chi )\approx( 70\%, 99\%)$ with $\br (Z' \to \nu \bar \nu)
\lesssim (3\% ,1\%)$ with a dark matter coupling $\schi = (0.84 ,
0.1)$.  For a $U(1)_{L_\mu-L_\tau}$ gauge boson, the decay into
neutrinos dominates even in the presence of dark matter. Both scale
with the gauge coupling $g_{Z'}$, and $\br (Z'\to \chi \bar
\chi)\approx (10\%, 20\%)$ for $\schi = (0.84 , 0.1)$. It is therefore
necessary to constrain the invisible $Z^\prime$ width to a similar
precision to either rule out or establish a link to dark
matter.\bigskip

As usual, invisible mediator decays lead to large missing transverse
energy in association with hard jets, Eq.\eqref{eq:lhc_signals}.  The
dominant backgrounds are $Z(\to\nu\nu)$+jets and $W(\to
l\nu)$+jets. The latter can be suppressed with a lepton veto, but a
fraction of events will remain if the lepton falls outside the
detector acceptance or does not meet the isolation requirements. Other
channels such as $t\bar t$ and $Z(\to ll)$+jets comprise less than 1\%
of the background and are not considered here.

We simulate the backgrounds with leading-order matrix elements, merged
with up to two additional jets in the parton shower using the CKKW-L
procedure, as implemented in \textsc{Sherpa}~\cite{sherpa}. For the
signal we rely on \textsc{MadGraph5}~\cite{madgraph} and
\textsc{Pythia8}~\cite{pythia}. Both, signal and background samples
are passed through the \textsc{Delphes}~\cite{delphes} detector
simulation with the ATLAS default detector card and $R=0.4$ anti-$k_T$
jets.  

As a start, we consider a standard cut-and-count analysis, following
an 8~TeV CMS analysis~\cite{cms_monojet}. We require a minimum
transverse energy $\met > 100$~GeV and a hard jet with $p_T > 100$~GeV
and $|\eta| < 2.5$. Events with a second jet only pass if $p_T >
30$~GeV, $|\eta| < 4.5$, and $\Delta\phi(j_1,j_2) < 2.5$, where the
last requirement suppresses QCD di-jets. Events with additional jets
with $p_T > 30$~GeV and $|\eta| < 4.5$ are vetoed, as are events with
one or more isolated leptons.

\begin{figure}[t]
\begin{center}
\includegraphics[width=\textwidth]{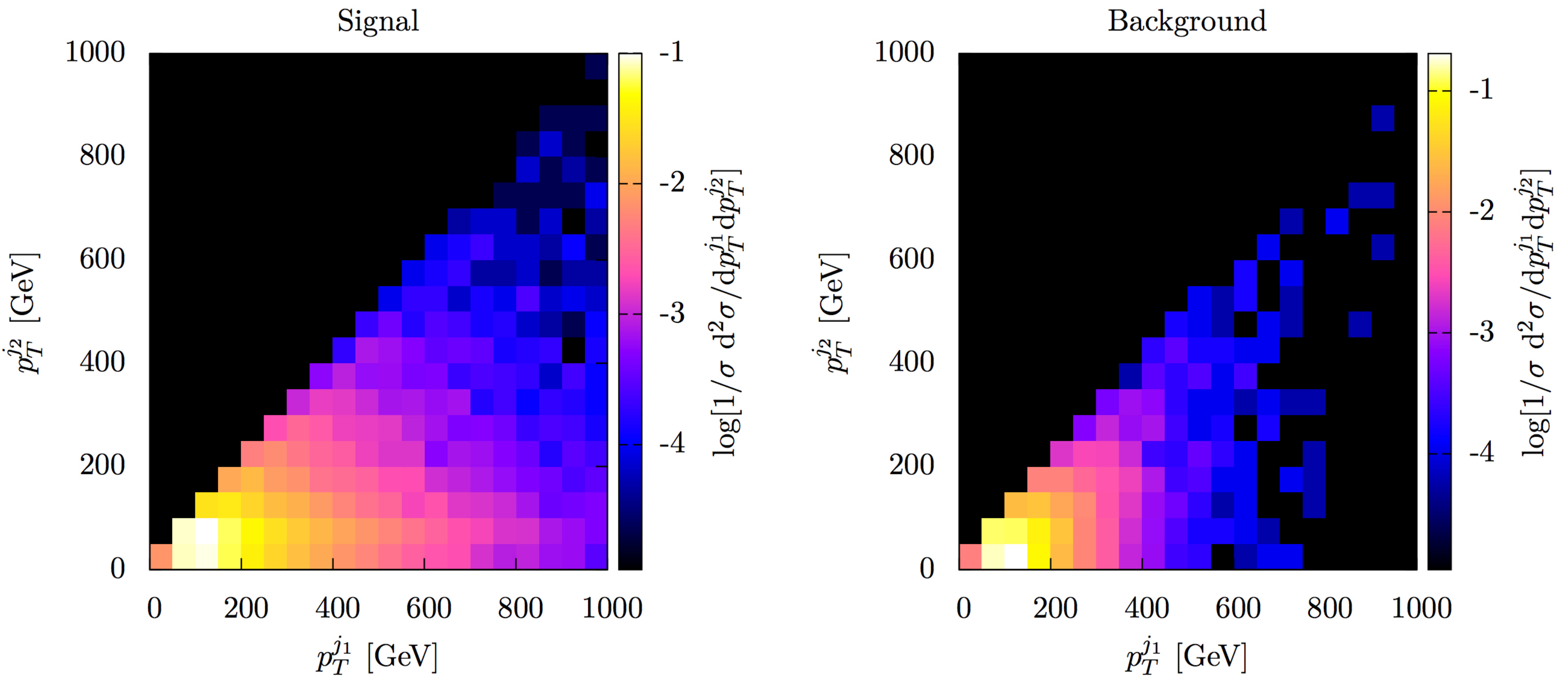}
\caption{\label{fig:2dhistos} The two-dimensional distributions of the leading and second jet $p_T$ for signal (left; same benchmark point as plotted above) and background (right).  }
\end{center}
\end{figure}

We select regions with $s/\sqrt{b+(\alpha s)^2+(\beta b)^2} > 2$,
where $\alpha$ and $\beta$ are systematic uncertainties on the signal
and background, respectively. The most excluded region is then used to
set the limit. The total signal rate is dominated  by the
low-$\met$ regime, with more than 80\% of signal events coming from
$\met < 400$~GeV for our model parameters. This implies that for
$300~\ifb$ our results are systematics limited, and it is instructive
to ask whether the precision can be improved by using the full $\met$
shape information of the $\met$ distribution. 

To this end we perform a binned likelihood analysis of the $\met$
distribution. Our procedure is based on the modified frequentist
$CL_s$ method~\cite{cls_method}. Further details, including the
modelling of systematics, can be found in the Appendix.  We highlight
the improvement over the standard approach in Fig.~\ref{fig:etmiss},
where we show the expected $CL_s$ limit in a currently allowed
parameter point as a function of the minimum $\met$ cut, both for a
shape analysis and for a cut-and-count analysis.  The limit from the
shape analysis gradually degrades as more bins are excluded and more
information is lost, while the cut-and-count limit moderately improves
when we apply a very stringent cut.  This shows how a simple counting
experiment above a stringent $\met$ cut is not the most effective way
of observing a mono-jet signal.
\bigskip

The choice of the $\met$ distributions can be further optimized by
including two-dimensional histograms, provided the proper correlations
between variables are available. For example, in
Fig.~\ref{fig:2dhistos} we show the correlation between the first and
second jet $p_T$, showing potential discriminating power. In practice,
including this information requires full control over the correlations
and a very large event sample to obtain a reliable estimate of the
event counts, so we merely comment that it is worth pursuing in
future.

\subsection{Exploiting $S$ decays}
\label{sec:lhc_scalar}

Our consistent model setup allows us to include the scalar mode in the
$Z'$ analysis. Given the observed relic density and the direct
detection constraints the scalar mass cannot be much larger than the
vector mass. Following Sec.~\ref{sec:constraints} the kinetic mixing
$\schi$ is strongly constrained, unless the $Z'$ is very heavy.  At
least in the $U(1)_{L_\mu-L_\tau}$ case the relic density can be
reproduced independently of $\schi$ through annihilation into leptons.
However, a sizable kinetic mixing is necessary to produce
the $Z'$, since any coupling between the $Z'$ and protons is
proportional to $\schi$ for both the $U(1)_X$ and
$U(1)_{L_\mu-L_\tau}$ models.
A simplified model with the $Z'$ mediator and a dark matter candidate
does not predict any relevant LHC signal.

In contrast to the kinetic mixing angle, the Higgs portal coupling
$\lambda_{HS}$ is not protected for example by an embedding in a
non-abelian gauge group. In the absence of an anomaly even without a
DM candidate, there is also no reason for the $S$ to couple to the
DM. This way the scalar mixing angle is not constrained by direct
detection and can be large.  This motivated searches for the vector
mediator in the process
\begin{align}
pp \to S \to Z'Z' \; ,
\end{align}
proportional to the scalar mixing angle $\sin\alpha$ and independent
of $\schi$.  Additional searches for $S\to Z Z'$ decays are possible,
but the corresponding partial width is again proportional to
$\schi$.\bigskip

\begin{figure}
\includegraphics[width=\textwidth]{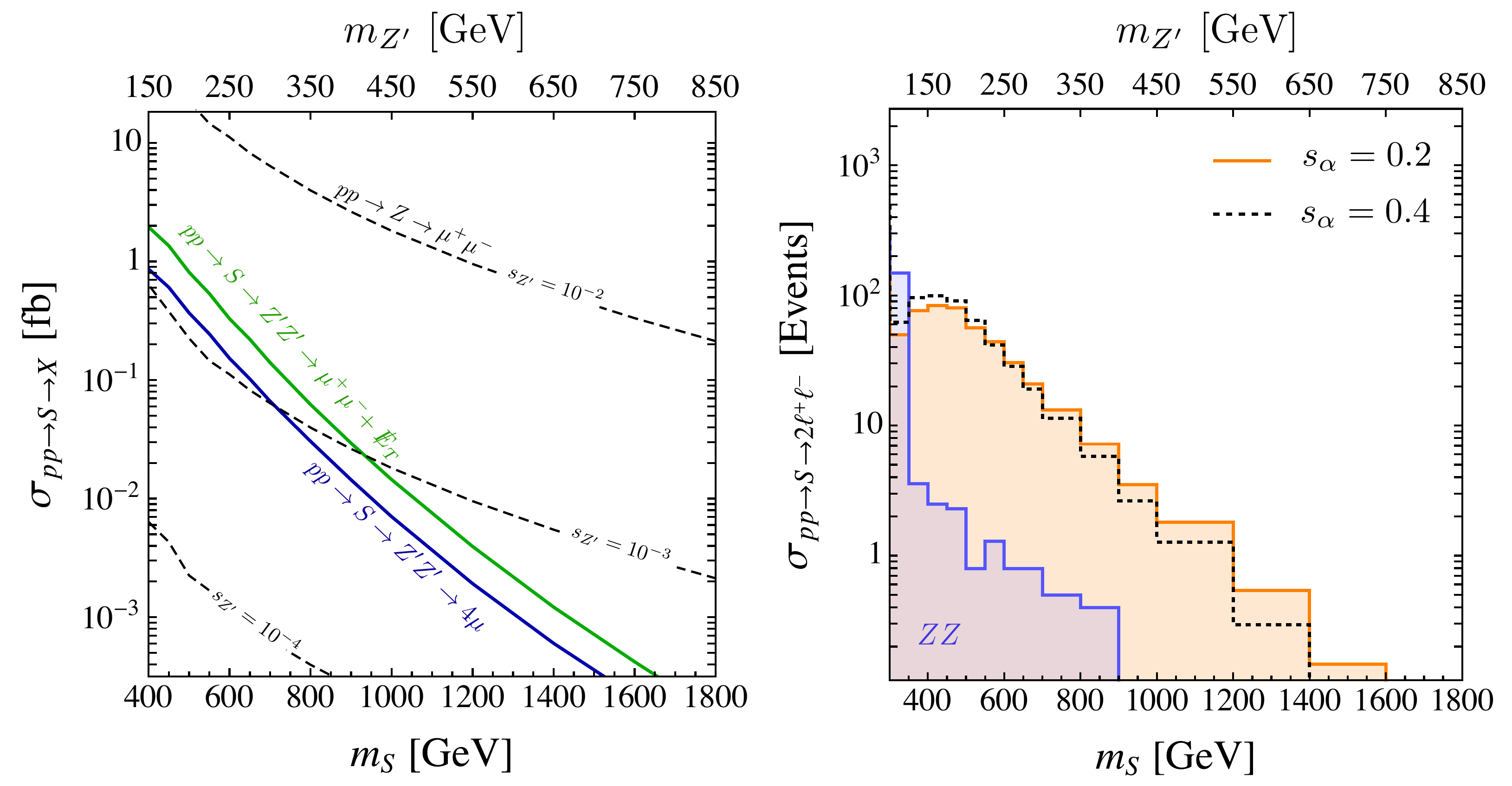}
\caption{Left: $S$-induced mono-$Z'$ and four-muon signal rates compared to
  the di-lepton resonance for $\schi=10^{-2}, 10^{-3}, 10^{-4}$.
  Right: signal and background events for $pp\to S \to 4\mu$
  assuming $s_\alpha=0.2$ and 0.4 after all cuts.}
\label{fig:crosssections} 
\end{figure}

The decay $S\to Z'Z'$ defines a mono-$Z'$ signal~\cite{ATLAS:2018fgp},
allowing for a discovery of a vector mediator through the scalar
portal. This signature is established for dark
radiation~\cite{Bai:2015nfa} and extended dark
sectors~\cite{Autran:2015mfa}. In consistent vector mediator models
the mono-$Z'$ signal is resonantly enhanced.  Another promising signal
is the competing decay
\begin{align}
S\to Z'Z'\to 4 \mu \; .
\end{align}
The two signals scale like
 \begin{align}
\frac{\sigma(pp\to S\to \ell^+\ell^- \met)}{\sigma(pp\to S\to 4 \ell)} 
\approx 
\frac{\Gamma(Z'\to \chi\bar \chi )}{\Gamma(Z'\to \ell^+\ell^-)}\; ,
\end{align}
with $\Gamma(Z'\to \chi\bar \chi )\propto g_{Z'}^2$. On the lepton
side, $\Gamma(Z'\to e^+e^-)\propto s_{\chi}^2$ for both $U(1)_X$ and
$U(1)_{L_\mu-L_\tau}$, while $\Gamma(Z'\to \mu^+\mu^-)\propto
s_{\chi}^2$ for $U(1)_X$ and $\Gamma(Z'\to \mu^+\mu^-)\propto
g_{Z'}^2$ for $U(1)_{L_\mu-L_\tau}$. A measurement of all three decays
would allow us to identify the underlying gauge group and constrain
the dark matter contribution to the invisible $Z'$ width.\bigskip

\begin{figure}
\includegraphics[width=\textwidth]{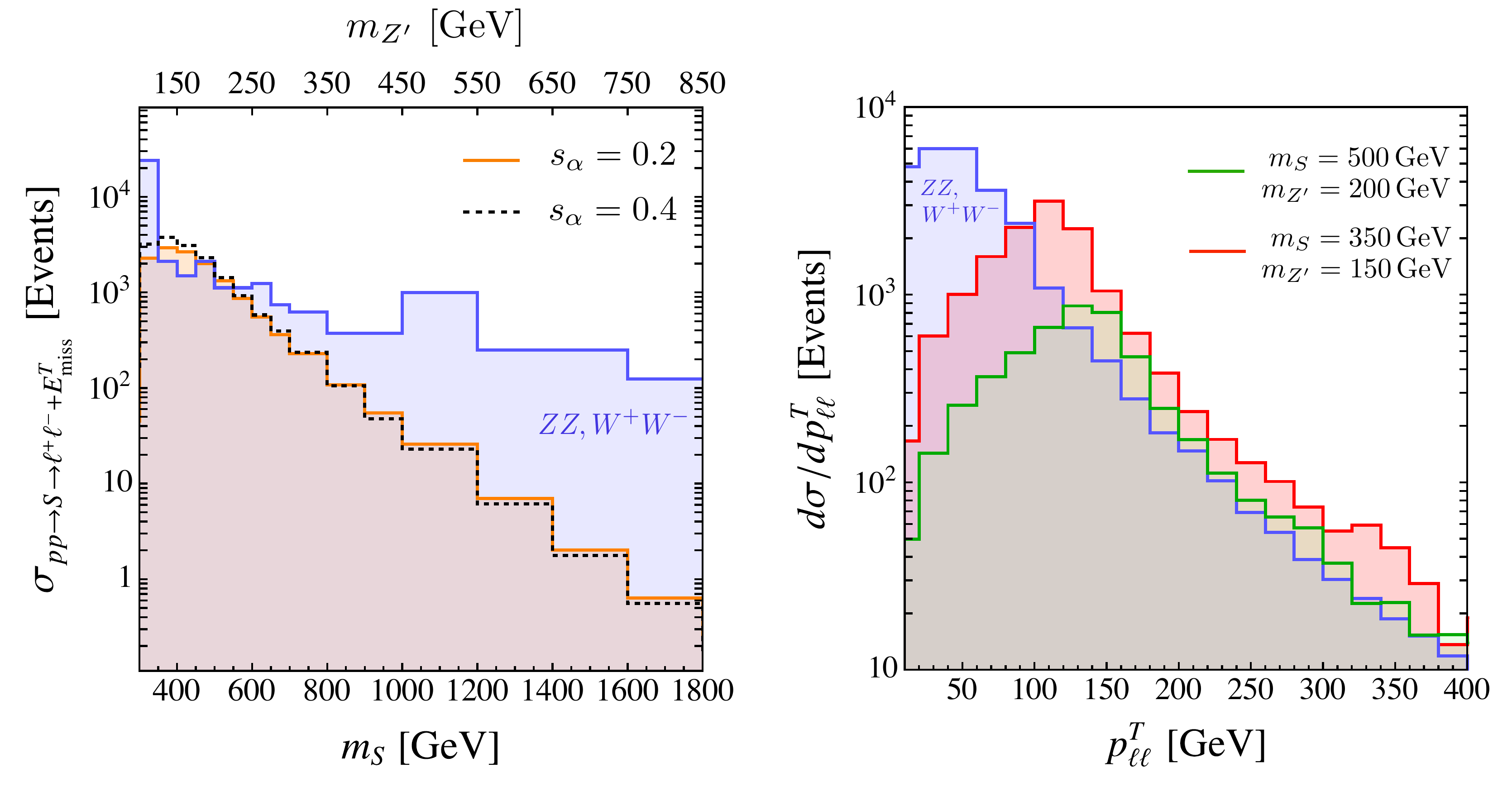}
\caption{Left: signal and background events for $pp\to S \to
  \mu^+\mu^- +\met$ assuming $s_\alpha=0.2$ and 0.4, after all
  cuts. Right: signal and background rates for two benchmark points.}
\label{fig:jacobian} 
\end{figure}

In the left panel of Fig.~\ref{fig:crosssections} we see how for $\sin
\alpha=0.4$, $m_S\lesssim 1.8$~TeV, and $s_{Z'} \sim 10^{-3}$, the
4-lepton and mono-${Z'}$ cross sections can exceed the di-lepton cross
section. We assume a collider energy of 14~TeV. In addition, the signal can be
easily extracted through the resonance conditions $m_{\ell\ell}\approx
m_{Z'}$ and $m_{Z'Z'}\approx m_S$. In the analysis we ask for two pairs of opposite sign muons reconstructing a $Z'$ each, and implement cuts on the invariant masses 
\begin{align}
m_{4\mu}=(1\pm 0.1)\, m_S
\qquad \text{and} \qquad m_{\mu\mu}=(1\pm 0.1)\, m_{Z'}\; , 
\end{align}
as well as $p_{T,\ell}> 20$ GeV for each muon. We show 
the $S \to 4 \mu$ signal and background rates
for an integrated luminosity of $3~\iab$, assuming $s_\alpha=0.2$ and
0.4 in the right panel of Fig.~\ref{fig:crosssections}. We fix the
gauge coupling to the maximum value $g_{Z'}=0.1-0.85$ allowed by the
indirect constraints in Sec.~\ref{sec:constraints}.  The blue contours
show the dominant $ZZ$ backgrounds after cuts. In the lowest mass bin, the overlap with the $Z$ resonance is responsible for the spike in background events. Smaller scalar mixing
angles do not necessarily result in fewer signal events once we take
into account the scaling of the decay widths $\Gamma(S\to
\text{SM})\propto s_\alpha^2$ and $\Gamma(S\to Z'Z')\propto
g_{Z'}^2$. An increased production rate is partially cancelled by a
reduced branching ratio $\br(S\to Z'Z')$.\bigskip

The 
mono-$Z'$ signal rate is larger than the 4-lepton rate 
by an order of magnitude throughout the parameter space. The
$p_{T,\ell\ell}$ spectrum of the signal displays a Jacobian peak
characteristic for the resonant decay. The maximal value
\begin{align}
p_{T, \ell\ell}^\text{max}\approx m_S\,\bigg(\frac{1}{4}-\frac{m_{Z'}^2}{m_S^2}\bigg)^{1/2}\,,
\end{align}
allows us to reduce the backgrounds through harder $\met$ cuts. We
show the $p_{T,\ell\ell}$ distribution for $(m_S=500,m_\chi= 200)$~GeV
and $(m_S=350,m_\chi= 150)$~GeV. We apply the cuts from
Ref.~\cite{CMS:2017gbj} and in addition require
\begin{align}
\met >100\,\gev 
\qquad \text{and} \qquad
p_{T,\ell\ell}> 
\begin{cases} 60 \,\gev & m_S < 600\, \gev \\
              100 \,\gev & m_S > 600\, \gev
\end{cases} \; .
\end{align}
The hardest lepton pair has to reconstruct the $Z'$ mass to $\pm
10\%$.  The signal and background are shown in the left panel of
Fig.~\ref{fig:jacobian} for $s_\alpha=0.2$ and 0.4 for different
masses $m_S$ and $m_{Z'}$ and gauge couplings $g_{Z'}=0.1 - 0.85$. Again, the overlap with the $Z$ resonance leads to the large number of background events in the first bin.
Even for a soft $\met$ cut the signal will be even more significant
than the 4-lepton signal because of the large signal rate.

In Fig.~\ref{fig:SoverB}, we show the significances of the
two $S$-induced signals for 
$s_\alpha= 0.2$ and 0.4.
For the small kinetic mixing angles implied by indirect
constraints and direct detection, the mono-$Z'$ signal can
be the discovery channel for a 
$U(1)_{L_\mu-L_\tau}$ mediator. Note that the results of this section
also hold for the gauge groups $U(1)_{L_e-L_\tau}$ and
$U(1)_{L_e-L_\mu}$ for $s_{Z'}\to 0$, taking into account the LEP
bounds of Eq.\eqref{eq:etaub}.

It is clear from Fig.~\ref{fig:SoverB} that a simple cut-and-count analysis offers little sensitivity above $m_s \simeq $1 TeV, even after applying cuts for an on-shell $Z^\prime$. Therefore, analogous to Sec. 6.2 we apply a shape analysis of the $p_{T,\ell\ell}$ spectrum shown in Fig.~\ref{fig:jacobian}\footnote{We do not perform the shape fit below $m_S <$ 900 GeV where the cut-and-count significance is already high enough to test the presence of a signal.}. We see a moderate gain from the shape analysis, since the distinctive Jacobian peak of the signal offsets the drop in sensitivity from the reduction in cross section, however the improvement is less substantial than in the mono-jet case, since the resonance cuts already suppress the background quite effectively.

\begin{figure}
\centering
\includegraphics[width=.5\textwidth]{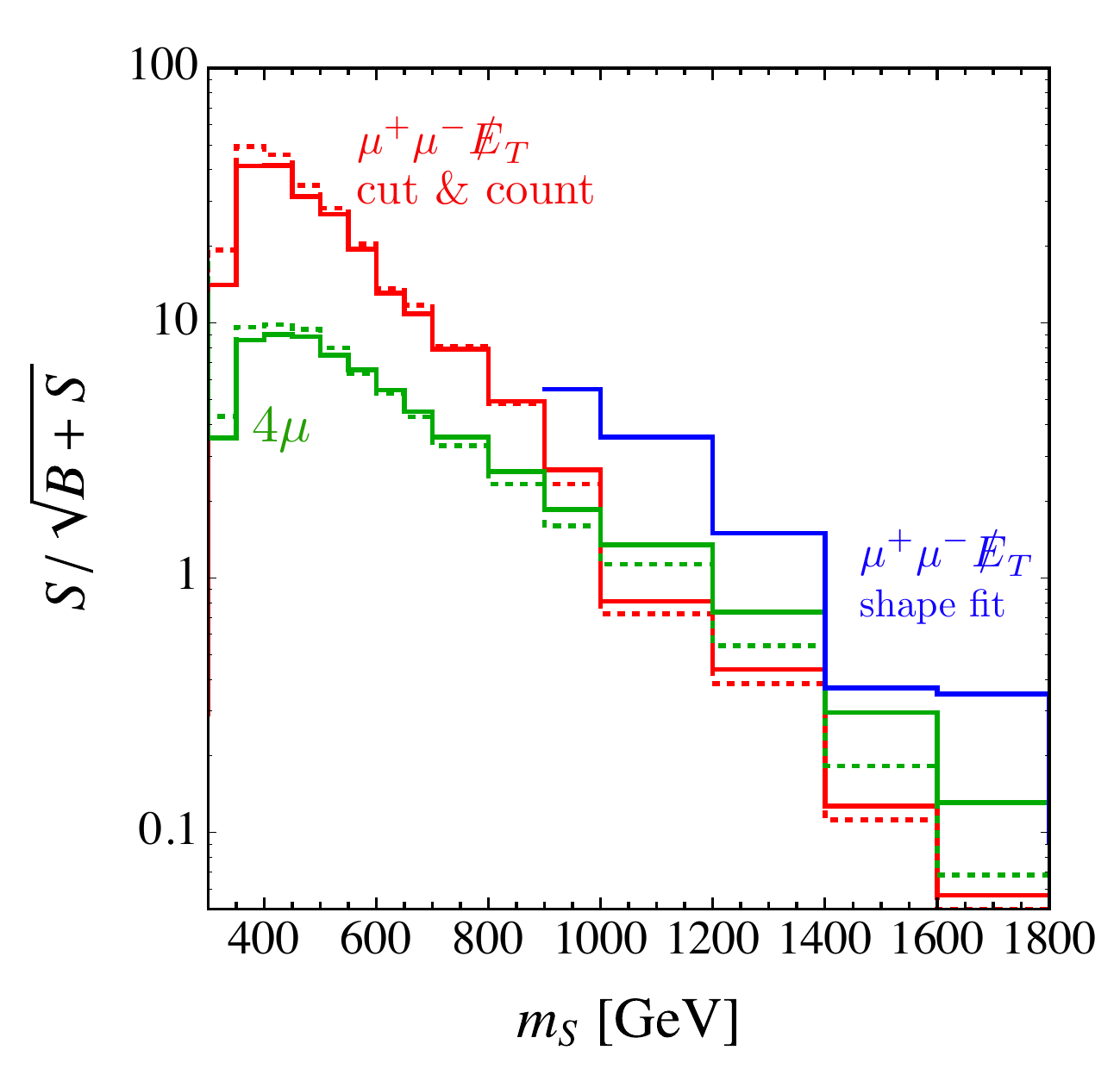}
\caption{Significance of the mono-$Z'$ signal in red in comparison to
  the 4-muon signal in green for difference scalar masses and for
  $s_\alpha=0.2 (0.4)$ shown by the solid (dotted) contours,
  respectively. The red (blue) contours correspond to the significance based on the cut \& count (shape fit) analysis.}
\label{fig:SoverB} 
\end{figure}

\section{Conclusions}

The best-motivated simplified models for dark matter with a vector
mediator are anomaly-free, gauged global symmetries of the SM. We
discuss several different such gauge groups, a $U(1)_X$ under which
only dark matter is charged and all couplings to the SM are mediated
through a kinetic mixing term, charged lepton family number
differences $U(1)_{L_e-L_\mu}$, $U(1)_{L_e-L_\tau}$ and
$U(1)_{L_\mu-L_\tau}$, and the gauged baryon-lepton number difference
$U(1)_{B-L}$. Obviously, mediators with tree-level couplings to
electrons are strongly disfavored by LEP bounds, leaving us with
$U(1)_X$ and $U(1)_{L_\mu-L_\tau}$ for a detailed study.

For the $U(1)_X$ model sizable kinetic mixing angles are necessary to
reproduce the observed relic density, which brings the model into
conflict with direct detection bounds. For the $U(1)_{L_\mu - L_\tau}$
mediator the relic density can be explained for sub-TeV masses and
order-one gauge couplings. Even allowing for loop-induced kinetic
mixing this parameter space is compatible with constraints from Planck
measurements of the CMB spectrum and direct detection. However, the
dark matter phenomenology constrains the mass splitting between the
vector mediator and the scalar mediator responsible for the $Z'$ mass
generation.
 
A common feature of the gauge groups we consider is a sizable
branching ratio $\br(Z'\to \nu \bar \nu)$. This introduces a mono-jet
signal even in the absence of a dark matter coupling. We discuss the
prospects of observing decays to dark matter by fitting the $Z'$-width
in the di-lepton channel and by precisely measuring the mono-jet
rate. In principle, the former is much more sensitive. However, for
$m_{Z'}\approx 1$~TeV the ATLAS and CMS energy resolution rule out
this method for $\Gamma_{Z'}< 5 (100)$ GeV for electrons (muons). In
this case a precise measurement of the mono-jet rate is indispensable
to establish mediator nature of the $Z'$ gauge boson. We explore the
additional sensitivity gained by a shape analysis of the $\met$
distribution compared to a cut-and-count analysis.

For the small kinetic mixing angles preferred by the dark matter
constraints, the $s$-channel production of the $Z'$ mediator at the LHC
is strongly suppressed. In contrast, the scalar mediator mode can be
produced through a Higgs portal. Since $U(1)_{L_\mu-L_\tau}$ is
anomaly free within the SM, the scalar does not have to couple to the
dark matter. Its dominant decay is $S\to Z'Z'$, if kinematically
allowed. The corresponding signatures are a resonantly enhanced
4-lepton signal $pp\to S \to Z'Z' \to 4\mu$ and a mono-$Z'$ signal
$pp\to S \to Z'Z' \to \mu^+\mu^- \met$. This combination is
characteristic for a consistent vector mediator model based on this
gauge group.  In particular the mono-$Z'$ final state with a leptonic
$Z'$ decay is a potential discovery channel for our consistent vector
mediator model.

\section*{Acknowledgments}

We thank Julian Heeck for useful comments regarding the structure of
the neutrino mass matrices and Dirk Zerwas for reminding us of the
correct lepton energy resolutions. 

\clearpage
\appendix
\section{Details of $U(1)$ extensions}\label{app:details}

The scalars in \eqref{eq:scalar} acquire VEVs
$\langle H\rangle=v/\sqrt{2}$ and $\langle S\rangle=v_S/\sqrt{2}$, and
the Higgs portal term induces the mixing
\begin{align}
\mathcal{M}_{H,S}^2= \begin{pmatrix}\lambda_H \,v^2& \lambda_{HS}\,v\,v_S\\
 \lambda_{HS}\,v\,v_S& \lambda_S \,v_S^2 
\end{pmatrix} \; .
\label{eq:matrix_scalars}
\end{align}
It can be diagonalized with a unitary rotation 
\begin{align}
\begin{pmatrix} S \\ H \end{pmatrix}
\to 
\begin{pmatrix} c_\alpha & s_\alpha \\
                -s_\alpha & c_\alpha \end{pmatrix} \; 
\begin{pmatrix} S \\ H \end{pmatrix}
\qqquad \text{with} \qquad 
t_{2\alpha}= \frac{2\lambda_{HS}\,v\,v_S}{\lambda_H\,v^2-\lambda_S\,v_S^2} \; ,
\end{align}
where $t_{2 \alpha} \equiv \tan (2 \alpha)$.\bigskip

The interaction with the SM-gauge sector allows for a
mixed kinetic term involving the Standard Model $U(1)_Y$-boson as given in \eqref{eq:kinmixlag},
where the notation $\hat B_{\mu\nu}$ indicates that the kinetic terms
of the gauge fields are not yet canonically normalized.  As indicated
by the above notation with $\schi \equiv \sin \theta_{Z'}$ we consider
kinetic mixing a phenomenon related to field rotations, but the term
$\schi$ in the Lagrangian does not arise from a rotation. Instead, it
is generally allowed by all symmetries at tree level and will
typically appear at one loop, even if it should vanish at tree
level. We assume $s_{Z'}<1$, otherwise the Lagrangian in
Eq.\eqref{eq:kinmixlag} corresponds to a theory with a single
propagating gauge boson ($s_{Z'}=1$) or a kinetic term with the wrong
sign ($s_{Z'}>1$).

For the abelian case the kinetic term can be diagonalized by an
orthogonal rotation in the two gauge fields. The problem with such an
orthogonal transformation is that it shifts the hypercharge and
eventually the electromagnetic current. To explicitly keep the
electromagnetic current and the canonical normalization, we introduce
a non-orthogonal rotation $G(\theta_{Z'})$ instead,
\begin{align}
\begin{pmatrix} \hat{B}_\mu \\ \hat{Z}'_\mu \end{pmatrix} 
= G(\theta_{Z'}) \, 
\begin{pmatrix} B_\mu \\ Z'_\mu \end{pmatrix}  
= \begin{pmatrix} 1  & -\schi/\cchi \\ 0 & 1/\cchi \end{pmatrix}
  \begin{pmatrix} B_\mu \\ Z'_\mu \end{pmatrix} \; .
\end{align}
Now the SM fermions couple
to the new gauge boson with a coupling strength
\begin{align}
j'_\mu\to \frac{1}{\cchi}\,j_\mu'-  \tchi \,j_\mu^Y \; ,
\end{align}
where $j_\mu^Y$ denotes the hypercharge current.
The combined mass matrix for the three electroweak gauge bosons
$B_\mu$, $W^3_\mu$, and $Z'_\mu$ reads
\begin{align}
\mathcal{M}_{B,W,Z'}^2
=\frac{v^2}{4}
\begin{pmatrix} g'^2  & -g\,g' &-{g'}^2 \tchi \\[2mm]
               -g\,g' & g^2    & g\,g' \,\tchi\\
               -{g'}^2 \,\tchi \quad  & \quad g\,g'\,\tchi \quad & \quad 2g_{Z'}^2\, \dfrac{q_S^2 v_S^2}{v^2 \cchi^2} + {g'}^2\, \tchi^2 
\end{pmatrix} \; ,
\label{eq:matrix_vectors}
\end{align}
where $g$ and $g'$ denote the $SU(2)_L$ and $U(1)_Y$ gauge couplings.
This mass matrix can be diagonalized through a combination of two
block-diagonal rotations with the weak mixing angle $\theta_w$ and an
additional angle $\theta_3$ in the lower-right block.  The mixing
angle $\theta_3$ is then given by
\begin{align}
\tan (2\theta_3)
&=
\frac{\sschi s_wv^2(g^2+{g'}^2)}{\cchi^2v^2(g^2+{g'}^2)(1-s_w^2\tchi^2)-2g_{Z'}^2q_S^2v_S^2} \notag \\
&= -\frac{\sschi s_w}{2g_{Z'}^2q_S^2}\frac{v^2}{v_S^2} \; \left( g^2+{g'}^2 \right)+\mathcal{O}\left( \frac{v^4}{v_S^4} \right) \; .
\end{align}
The physical gauge boson masses
\begin{align}
m_\gamma &= 0 \notag \\
m_{Z,Z'}^2&=
\frac{1}{8 \cchi^2} \bigg[\cchi^2v^2(g^2+{g'}^2)+{g'}^2\schi^2v^2+2g_{Z'}^2q_S^2 v_S^2\notag \\
&\quad\pm\sqrt{\left( \cchi^2v^2(g^2+{g'}^2)+{g'}^2\schi^2v^2+2g_{Z'}^2q_S^2v_S^2\right)^2 
              +8\cchi^2g_{Z'}^2q_S^2v^2v_S^2(g^2+{g'}^2)} \bigg] \notag \\
&=\begin{cases} 
 \dfrac{v^2}{4}(g^2+{g'}^2) \; \left(1-\dfrac{v^2}{v_S^2} \; \dfrac{\schi^2{g'}^2}{8g_{Z'}^2 q_S^2}\right)
 + \mathcal{O}\left( \dfrac{v^6}{v_{S}^4} \right) \\[.4cm]
   \dfrac{g_{Z'}^2q_S^2v_S^2}{2\cchi^2} + \dfrac{v^2}{4}{g'}^2\tchi^2
 + \mathcal{O}\left( \dfrac{v^4}{v_S^2} \right)  \; .
       \end{cases}
\end{align}
We show approximate results for $v_S>v$, motivated by our expectation
$m_{Z'}, m_S > m_Z$. The alternative series in terms of a small mixing
angle $\schi$ would have to be motivated by specific model
considerations~\cite{patrick}. \bigskip

A combination of all three rotations by the kinetic mixing parameter
and the angles $\theta_w$, $\theta_{Z'}$, and $\theta_3$ appears in the couplings of the
fermionic currents to the boson mass eigenstates,
\begin{align}
\left(ej_\text{em} , \frac{ej_Z}{s_w c_w}, g_{Z'}j_{Z'}\right) 
\begin{pmatrix}\hat A\\ \hat Z\\\hat Z'\end{pmatrix}
=&\left(ej_\text{em} , \frac{e}{s_w c_w} j_Z, g_{Z'}j_{Z'}\right) \,K\,\begin{pmatrix}A\\ Z\\  Z'\end{pmatrix} \notag \\
K=&\left[ R_1(\theta_3)R_2(\theta_w)G^{-1}(\theta_{Z'})R_2(\theta_w)^{-1}\right]^{-1} \notag \\
 =& \begin{pmatrix}
1 & -c_w s_3 \,\tchi & -c_w c_3 \,\tchi\\
0 & c_3 +s_w s_3 \tchi & c_3 s_w \tchi-s_3 \\
0 & s_3/\cchi &  c_3/\cchi 
\end{pmatrix} \; .
\label{eq:all_mixings}
\end{align}
The interesting aspect is that the combination of all angles is not an
orthogonal rotation. This is why the electromagnetic fermion current
of SM fermions couples to all three gauge bosons.\bigskip

Similarly, the complex mixing pattern affects the otherwise simple
coupling structure of the gauge boson to the two scalars
 \begin{align}
\begin{pmatrix} A& Z & Z'\end{pmatrix} \begin{pmatrix}0&0&0\\
 0&\multicolumn{2}{c}{\raisebox{-.5\normalbaselineskip}[0pt][0pt]{$W$}}\\
 0&&\end{pmatrix} \begin{pmatrix} A\\ Z\\ Z'\end{pmatrix} \; ,
\end{align}
with the massive sub-matrix 
\begin{align}
W&=-\frac{v s_\alpha}{8}\begin{pmatrix}
(g^2+{g'}^2) & (g^2+{g'}^2) s_w \tchi \notag \\[6pt]
(g^2+{g'}^2) s_w \tchi & \quad (g^2+{g'}^2) \tchi^2s_w^2 - \dfrac{4 g_{Z'}^2 q_S^2}{t_\alpha \cchi^2} \dfrac{v_S}{v}
\end{pmatrix}
S\\
&\quad  +\frac{v c_\alpha}{8}\begin{pmatrix}
(g^2+{g'}^2) & (g^2+{g'}^2) s_w \tchi \\[6pt]
(g^2+{g'}^2) s_w \tchi & \quad (g^2+{g'}^2) \tchi^2s_w^2 + \dfrac{4 g_{Z'}^2q_S^2 t_\alpha}{\cchi^2} \dfrac{v_S}{v}
\end{pmatrix} H
+\mathcal{O} \left( \frac{v^2}{v_S} \right) \; .
\end{align}
This matrix induces new couplings between the scalars $H$ or $S$ and
the gauge bosons $Z$ and $Z'$. They follow a generic hierarchy of
couplings
\begin{align}
\frac{g_{SZZ'}}{g_{HZZ'}} \propto t_\alpha\approx \frac{1}{3} \; ,
\label{eq:coup_ratio}
\end{align}
because the scalar mixing angle is constrained by Higgs coupling
strength measurements $\sin\alpha < 0.3$~\cite{legacy}.\bigskip

It is instructive to link those three gauge groups to neutrino
masses~\cite{mu_tau_zprime_neut}. For gauged $U(1)_{L_i-L_j}$
symmetries the three lepton generation carry different charges, which
implies that the leptons cannot mix and the Yukawa matrix is
diagonal. The same is true for the neutrinos, once we add right-handed
neutrinos only charged under the new gauge group. The right-handed
neutrinos also have a Majorana mass. For example in the case of
$U(1)_{L_\mu-L_\tau}$ such a Majorana mass term can appear as the
$(e,e)$ entry and in the $(\mu,\tau)$ and $(\tau,\mu)$ entries. In
addition, terms of the kind $yNNS$ lead to Majorana masses when the
new scalar is replaces by its VEV. Still, $S$ is charged under the new
$U(1)$ group, which leads to possible $(e,\mu)$ and $(e,\tau)$
entries. The corresponding, symmetric Majorana mass matrix for three
generations of neutrinos reads
\begin{align}
\begin{pmatrix} m_e & y_{e,\mu} v_S & y_{e,\tau} v_S \\
y_{e,\mu} v_S & 0 &m_{\mu,\tau} \\
y_{e,\tau} v_S & m_{\mu,\tau} & 0
\end{pmatrix} \; ,
\end{align}
assuming 
\begin{align}
q_S = 1 \; .
\label{eq:qs_one}
\end{align}
As a consequence of the diagonal mass matrices for the charged
leptons, the $Z'$ gauge boson has no lepton-flavor violating couplings
to charged leptons and flavor-changing neutral currents only arise at
the one-loop level.  From this construction it is clear that the
generation-universal groups $U(1)_X$ and $U(1)_{B-L}$ do not have this
direct link to neutrino masses.

\section{Mono-jet shape analysis}

A shape analysis like the one discussed in Sec.~\ref{sec:lhc_inv}
typically distinguishes a background-only hypotheses $H_0$ from a
signal-plus-background hypothesis $H_1$. The Neyman-Pearson lemma
states that the most powerful test statistic is the likelihood ratio.
For a counting experiment in the absence of systematic uncertainties
it is given by Poisson probabilities for obtaining $d$ data events
given the expectation values $s+b$ and $b$.  In practice, we usually
take its logarithm,
\begin{align}
-2 \log Q = -2 \log \frac{ P(d | s+b)}{P(d | b)} 
= -s + d \log \frac{s+b}{b} \; .
\label{eqn:poiss} 
\end{align}
In this form we can easily combine different channels of bins of a
distribution and therefore perform a shape analysis for example of a
$\met$ distribution.\bigskip

To compute confidence levels we numerically evaluate the corresponding
$p$-values by generating a large number of Monte Carlo
pseudo-experiments, with $CL_{s+b}$ being the fraction of
pseudo-experiments that generate at least as many events as observed
in the data.  Instead of excluding regions for which $CL_{s+b} \le
0.05$, we take the $CL_s$ procedure~\cite{cls_method}, which only
excludes this hypothesis if $CL_{s+b}/(1-CL_b) \le 0.05$. This is more
robust against spuriously high sensitivity when both $s$ and $b$ are
small, at the price of being conservative otherwise.

One way of including systematic uncertainties is by convoluting the
individual Poisson likelihoods in Eq.~\ref{eqn:poiss} with Gaussians.
This procedure reduces the sensitivity by smearing the log-likelihood
distributions for the two hypotheses, thus reducing the distinction
between $s$ and $s+b$.

Clearly, the separation between the hypotheses and thus the final
confidence level is extremely sensitive to the modelling of systematic
uncertainties. Therefore it is crucial to correctly propagate
systematics in the limit-setting procedure when using the full shape
information from binned distributions.  We study four scenarios, in
order of increasing conservatism: (i) no systematics at all; (ii)
uncorrelated bin-by-bin systematics; (iii) a 5\% correlation between
each bin and its nearest neighbor with all other correlations zero;
and (iv) a flat systematic fully correlated across all bins.\bigskip

\begin{figure}[t]
\begin{center}
\includegraphics[width=\textwidth]{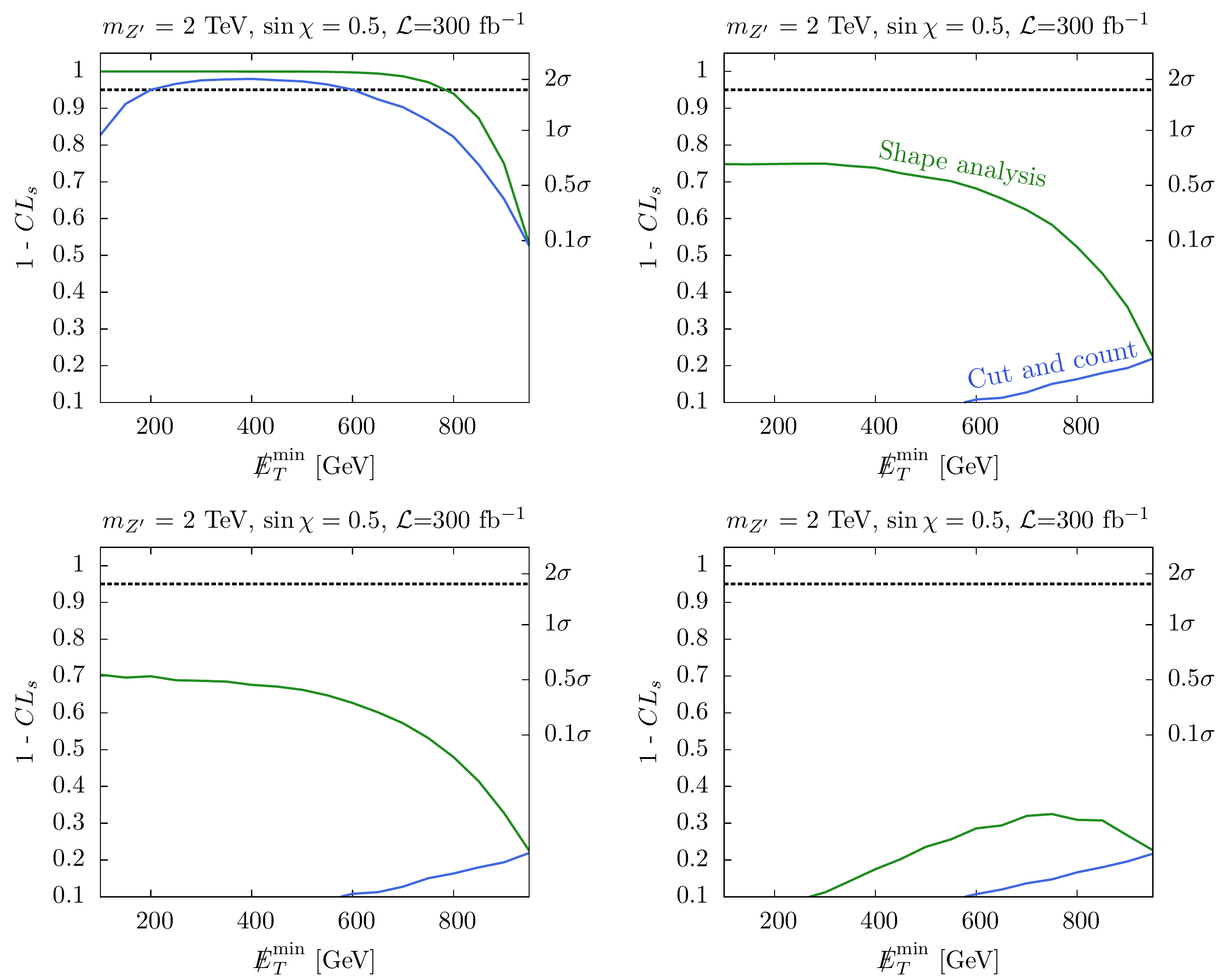}
\end{center}
\caption{Expected $CL_s$ for excluding the signal hypothesis from a
  $\met$ distribution in mono-jet events as a function of minimum
  $\slashed{E}_T$. We show the results from a full shape analysis
  (green) vs counting all events above the cut as a single bin
  (blue). Four systematics scenarios are considered: no systematics
  (top-left), an uncorrelated 5\% per-bin background uncertainty
  (top-right), a 5\% per-bin background uncertainty plus 100\%
  correlation between neighbouring bins (bottom-left), and a 5\%
  uncertainty fully correlated across all bins (bottom-right).}
\label{fig:smear}
\end{figure} 

As input data we use the binned mono-jet $\slashed{E}_T$ distributions
for the signal and the combined $Z$+jets and $W$+jets background for
$300~\ifb$ of data. As benchmark point for the test hypothesis, we
consider the $U(1)_X$ model discussed in Sec.~\ref{sec:lhc_inv} for a
$Z^\prime$ mass of 2~TeV and mixing angle $\sin\chi$ = 0.5. In
Fig.~\ref{fig:smear} we show $CL_s$ as a function of a minimum $\met$
cut for each of the four systematics scenarios, both using the full
shape information and using the integrated rate only (cut and count).

Beginning with the unrealistic case of no systematics we see that the
full shape analysis provides much more sensitivity than the
cut-and-count analysis in the low $\met$ region, reflecting the much
larger background there. For an uncorrelated 5\% systematic on the
background in each bin we see a lower significance for both shape and
rate analyses, but using shape information carries much better
discriminating power than cutting on $\met$ and counting events.

To estimate the effects of bin migration, we then include a full
correlation between neighbouring bins, with all other correlation
coefficients set to zero.  This has a mild influence on the
significance from the shape analysis, but does not affect our
conclusion that the full shape information is a more powerful
discriminator. Finally, we consider the extreme scenario of full
correlations across all bins. Adding more bins below $\sim$ 700~GeV
now leads to less discriminating power, because the 5\% uncertainty on
the background in the low-$\met$ region is smeared across all
bins. The behavior turns over around $\met = 700$~GeV, where
statistics becomes the main driver of discriminating power.

\end{fmffile}

\clearpage

\end{document}